\documentclass[12pt,english]{article}
\usepackage{amsmath,amssymb}
\usepackage{epsfig}
\usepackage{booktabs}
\usepackage{datetime}
\usepackage{cite}
\usepackage{color}
\usepackage{babel}
\usepackage{verbatim}
\usepackage{amstext}
\usepackage{graphicx}
\usepackage{esint}
\usepackage[unicode=true,
 bookmarks=true,bookmarksnumbered=false,bookmarksopen=false,
 breaklinks=true,pdfborder={0 0 1},backref=false,colorlinks=true]
 {hyperref}
\hypersetup{pdftitle={Title},
 pdfauthor={Authors},
 pdfsubject={Subject},
 pdfkeywords={Keys}} 
 
\def\m@th{\mathsurround=0pt }
\def\leftrightarrowfill{$\m@th \mathord\leftarrow \mkern-6mu
	\cleaders\hbox{$\mkern-2mu \mathord- \mkern-2mu$}\hfill
	\mkern-6mu \mathord\rightarrow$}

\def\overleftrightarrow#1{\vbox{\ialign{##\crcr
	\leftrightarrowfill\crcr\noalign{\kern-1pt\nointerlineskip}
	$\hfil\displaystyle{#1}\hfil$\crcr}}}

\newcommand{\be}{\begin{equation}}
\newcommand{\ee}{\end{equation}}

\def\I{\rm 1\kern-.24em l}  % Yes, I know. This ain't capital.

\def\shat{\ifmmode \hat{s}\else $\hat{s}$\fi}

\def\mc{\mathcal}
\def\ord{{\mathcal O}}

\def\nn{\nonumber}

\newcommand{\newc}{\newcommand}

\newc{\gsim}{\lower.7ex\hbox{$\;\stackrel{\textstyle>}{\sim}\;$}}
\newc{\lsim}{\lower.7ex\hbox{$\;\stackrel{\textstyle<}{\sim}\;$}}
\newc{\ie}{{\it i.e.}}
\newc{\etal}{{\it et al.}}
\newc{\mev}{\hbox{\rm\,MeV}}
\newc{\gev}{\hbox{\rm\,GeV}}
\newc{\tev}{\hbox{\rm\,TeV}}
\newc{\xpb}{\hbox{\rm\, pb}}
\newc{\xfb}{\hbox{\rm\, fb}}

\newc{\h}{{\cal H}}
\newc{\D}{{\cal D}}
\newc{\E}{{\cal E}}
\newc{\x}{{\widehat x}}
\newc{\q}{{\widehat q}}

\newc{\mtop}{m_t}
\newc{\mbot}{m_b}
\newc{\mz}{M_Z}
\newc{\mw}{M_W}
\newc{\alphasmz}{\alpha_s(M_Z)}
\newc{\swsq}{\sin^2\theta_W}
\newc{\cwsq}{\cos^2\theta_W}
\newc{\tw}{\tan\theta_W}
\newc{\cw}{\cos\theta_W}
\newc{\sw}{\sin\theta_W}
\newc{\BR}{\hbox{\rm BR}}
\newc{\zbb}{Z\to b\bar}
\newc{\Gb}{\Gamma (Z\to b\bar b)}
\newc{\Gh}{\Gamma (Z\to \hbox{\rm hadrons})}
\newc{\sgn}{\mbox{sgn}}

\def\I{1\hspace{-4pt}1}

\newcounter{mysubequation}[equation]

\def\beq{\begin{equation}}
\def\eeq{\end{equation}}
\def\bea{\begin{eqnarray}}
\def\eea{\end{eqnarray}}
\def\slashchar#1{\setbox0=\hbox{$#1$}           % set a box for #1
   \dimen0=\wd0                                 % and get its size
   \setbox1=\hbox{/} \dimen1=\wd1               % get size of /
   \ifdim\dimen0>\dimen1                        % #1 is bigger
      \rlap{\hbox to \dimen0{\hfil/\hfil}}      % so center / in box
      #1                                        % and print #1
   \else                                        % / is bigger
      \rlap{\hbox to \dimen1{\hfil$#1$\hfil}}   % so center #1
      /                                         % and print /
   \fi}                                         %
\catcode`@=11
\long\def\@caption#1[#2]#3{\par\addcontentsline{\csname
  ext@#1\endcsname}{#1}{\protect\numberline{\csname
  the#1\endcsname}{\ignorespaces #2}}\begingroup
    \small
    \@parboxrestore
    \@makecaption{\csname fnum@#1\endcsname}{\ignorespaces #3}\par
  \endgroup}
\catcode`@=12

\begin{document}

\baselineskip=18pt

\setcounter{footnote}{0}
\setcounter{figure}{0}
\setcounter{table}{0}

\begin{titlepage}
\begin{flushright}
%??-??-...
\end{flushright}
\vspace{.3in}
\hfill PP-012-005
\vspace{2.0cm}

\begin{center}
{\Large \bf
Probing the Scattering of \\ Equivalent  
Electroweak Bosons
}
\\[6pt]
\vspace{1.0cm}
{\bf Pascal Borel$^{a}$, Roberto Franceschini$^{a,b}$, Riccardo Rattazzi$^{a}$ and Andrea Wulzer$^{c,d}$}

\vspace{0.5cm}

\centerline{$^{a}${\it Institut de Th\'eorie des Ph\'enom\`enes Physiques, EPFL,
  CH--1015 Lausanne, Switzerland}}
  \centerline{$^{b}${\it Department of Physics, University of Maryland, College Park, MD 20742, U.S.A.}}
\centerline{$^{c}${\it Dipartimento di Fisica e Astronomia and INFN, Sezione di Padova, }}
\centerline{{\it Via Marzolo 8, I-35131 Padova, Italy}}
\centerline{$^{d}${\it Institute for Theoretical Physics, ETH Zurich,
8093 Zurich, Switzerland}}

\end{center}
\vspace{.8cm}

\begin{abstract}
\medskip
\noindent
We analyze the kinematic conditions under which the scattering of equivalent massive spin-1 vector bosons
factorizes out of the complete process.
In practice, we derive the conditions for the validity of the effective  $W$ approximation, proposed long ago but never established on a firm basis. We also present a parametric estimate of the corrections to the approximation and explicitly check its validity  in two examples. 
\end{abstract}

\end{titlepage}

\section{Introduction}
High-energy scattering  among longitudinally polarized electroweak vector bosons ($W_L$) is obviously the most direct
probe of the dynamics of electroweak symmetry breaking (EWSB) \cite{Chanowitz}. If a light Higgs boson ($H$) exists, then 
processes involving $H$, along with $W_L$, are similarly important \cite{Contino:2010mh}. 
 Unfortunately $W$ beams are not available,
and so we cannot, strictly speaking, directly study their collisions. However that should not be a problem of principle,
as we know very well that, because of initial state radiation, the quanta that probe short distance physics in high energy collisions are all eminently virtual. So, for instance, even though quarks and gluons do not even exist as asymptotic states, we can still study their short distance interactions via the parton model, or, more formally, thanks to factorization theorems. With the intuition offered by the parton model,  it is then natural to expect that at sufficiently high energy, much above the $W$ mass $m$, and with a proper selection of the kinematics of the final states, one should also be able to effectively test the collisions of $W$ bosons. Considering the prototypical reaction $q {\overline{q}}\to q'{\overline{q}}' WW$ one would expect that to be the case in the regime where the transverse momentum $P^W_\bot$ of the $W$'s is much larger than that of the final state fermions $p_\bot$, that is where the latter are forwardly emitted. In that regime it seems intuitively obvious that the class of diagrams shown in Fig.~1a will dominate over all others, like for instance Fig.~1b, because the low virtuality $V_1$ and $V_2$ of the intermediate $W$, which characterizes the forward emission, parametrically enhances their propagators. Furthermore, in first approximation one expects to be able to neglect $V_{1,2}$ in the hard  $WW\to WW$ subamplitude and to replace it with that for  {\it equivalent} on-shell quanta.  The differential cross section would then nicely factorize into a term describing $q\to q'W$ splitting, which is controlled by known interactions,  times the hard $WW\to WW$ scattering which is where new phenomena may well appear. This idea, inspired by the equivalent photon method \cite{equivalentphoton}, is  known as the effective W approximation (EWA) \cite{Dawson:1984gx,EWAbulk} and was largely employed in the  80's to simplify and speed up the computations of processes involving $WW\to X$. However, the conceptual validity of EWA 
was questioned by various authors, both in the early days and also more recently \cite{Kleiss:1986xp,hep-ph/0608019}. It is only in the case of a heavy Higgs, with strongly-coupled EWSB, that the issue has been set firmly, in favor of the validity of EWA (and factorization), in a very nice paper by Kunszt and Soper (KS) \cite{Kunszt:1987tk} in the late '80's. Later, the advent of powerful packages that allow to reliably and easily compute the exact cross section \cite{compute0l}, including also radiative corrections \cite{compute1l}, has seemingly made EWA obsolete. Indeed, one often heard  remark is that there is no way to single out diagrams with the scattering topology of Fig.~1a, and that all diagrams are equally important. We find that viewpoint disturbing for at least two reasons. On one side because it seems to entail that factorization fails for massive vector particles. On the other, because it suggests that it simply does not make sense, even in an ideal experimental situation, to extract in a model independent way the on-shell $\langle WWWW\rangle$ correlator from experimental data: the interesting physics of $WW$ scattering would always be mixed up in an intricate way with SM effects. We thus believe that studying the conditions for the applicability of EWA is important, and  timely as well. Obviously the goal is  not to find a fast and clever way to do computations. One should view  EWA as a selection tool that allows to identify the relevant 
kinematic region of the complete process, the one which is more sensitive to the EWSB dynamics. One would want  to focus on the kinematics where EWA applies not to speed up the computations, but to gain sensitivity to the relevant physics. 
 
 %%%%%%%%%%%%%%%%%%%%%%%%%%%%%%%%%%%%%
\begin{figure}
\begin{center}
\includegraphics[height=0.3\linewidth]{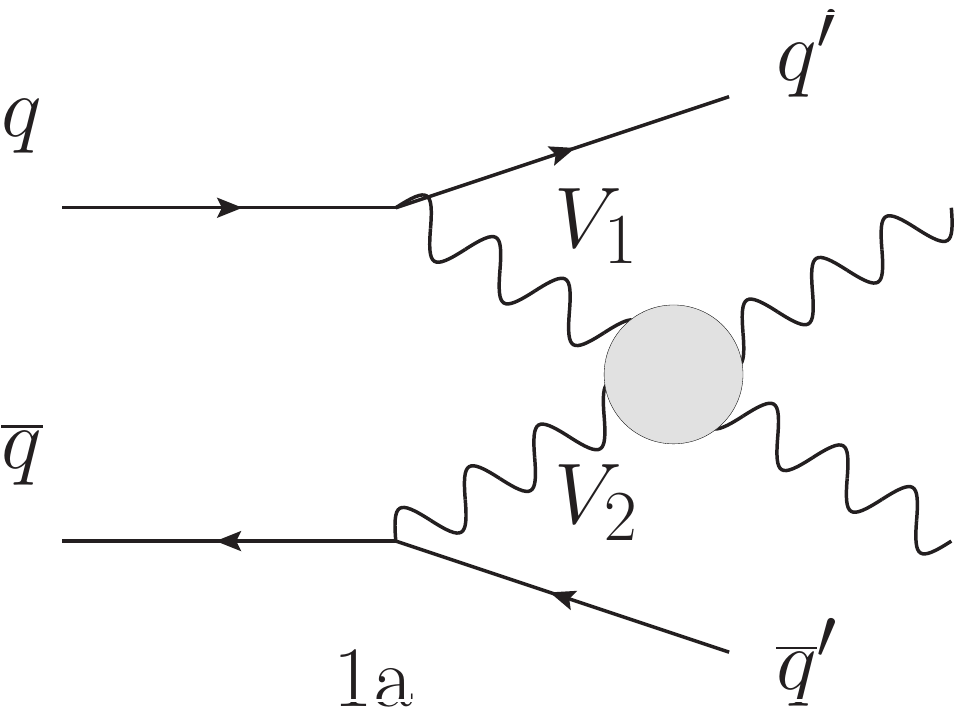}
\hspace{30pt}
\includegraphics[height=0.3\linewidth]{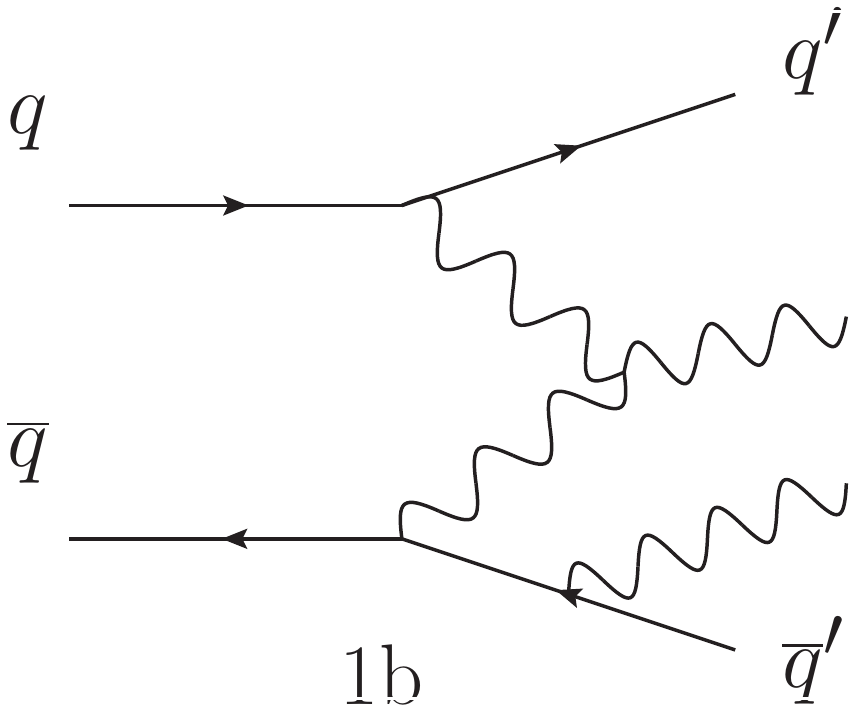}
\end{center}
\caption{Diagrammatic contributions to the $q {\overline{q}} \to q' {\overline{q}}' WW$ process. On the left, the scattering topology. On the right, one representative ``radiation'' diagram.}
\label{figintro}
\end{figure}
%%%%%%%%%%%%%%%%%%%%%%%%%%%%%
 
In this paper we shall analyze in  detail the applicability of EWA. We will find, not surprisingly, that, in the proper kinematic regime, factorization is valid and  EWA works egregiously. In order to prove that, we shall not need to focus, as KS did, on the case of a heavy Higgs or a strongly interacting EWSB sector, actually we shall not even need to restrict on the specific sub-process $WW\to WW$. Factorization indeed does not rely in any way on the detailed nature of the hard sub-process. It relies instead on the existence of a large separation of virtuality scales between the sub-process and the collinear $W$ emission. That only depends on kinematics and corresponds to requiring forward energetic jets and hard high $P_\bot$ outgoing $W$'s. When those conditions are imposed  EWA works well, for both longitudinally and transversely polarized $W$'s, also including the case of weakly-coupled  EWSB  (light and elementary Higgs) where all helicities interact with the same strength $\sim g_W$ at all energies. 

 One serious issue in the applicability of EWA is the size of the subleading corrections. In practice one would like to know how well it can be applied at the LHC with 14 TeV in the center of mass. A detailed quantitative answer to this question is beyond the scope of the present paper, we shall content ourselves with deducing a parametric estimate of the corrections. Our result is that typically the corrections scale quadratically with the inverse of the hardness  $H$ of the $WW$ collision, and can be estimated as $\sim m^2/H^2$ and $p_\bot^2/H^2$~\footnote{Parametric enhancements are possible but only in very peculiar situations, as we shall discuss.}. We will give a numerical confirmation of this scaling, finding however a typically large numerical prefactor that enhances the corrections. 

The paper is organized as follows. In section~2.1 we will introduce the EWA formula, discuss the basic physical reasons for its validity and state our results concerning the scaling of the corrections. In section~2.2 we will confirm these results in two examples, the processes \mbox{$q W^-\to q' W^+W^-$} and \mbox{$q {\overline{q}} \to q' \overline{q}' W^+W^-$}, by comparing explicitly the exact cross section with its EWA approximation. Section~3 contains our analytical derivations, and constitutes the main part of the paper. In section~3.1 we introduce the technical tools that are needed in our derivation; in section 3.2 we compute the amplitude for the ``soft'' $q\to q'W$ splitting and, finally,  in section~3.3 we  derive the EWA formula. We discuss the parametric estimate of the corrections in section~3.4, while section~3.5  illustrates some aspects of our derivation in the  case of the $WW\to WW$ sub-process. Finally, we report our conclusions in section~4.

\section{Basic Picture and Explicit Checks}

While the general derivation of the EWA, which 
we will present in section~3, is rather technical, the basic picture that underlies its validity is instead 
very simple and intuitive. In the present section we will first of all illustrate this picture and 
afterwards we will check the approximation 
explicitly in the relevant example of $WW\rightarrow WW$ scattering 
in the SM with a light Higgs. By the numerical comparison among the exact and the EWA cross sections we will also establish the scaling of the corrections to the approximation, a scaling which we will deduce analytically and cross-check in section~3.

\subsection{Basic Picture \label{BasicPicture}}

%%%%%%%%%%%%%%%%%%%%%%%%%%%%%%%%%%%%%
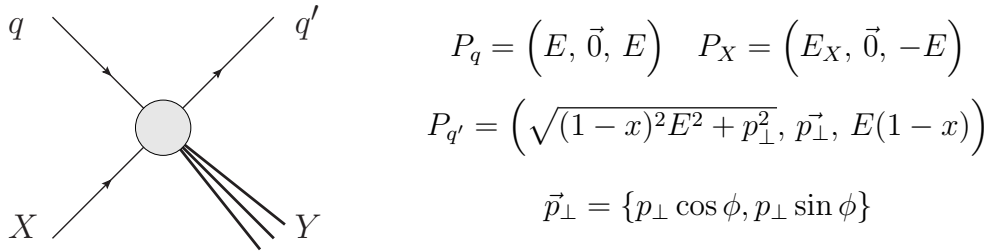
\begin{figure}
\hspace{0.12\linewidth}
\includegraphics[width=0.26\linewidth]{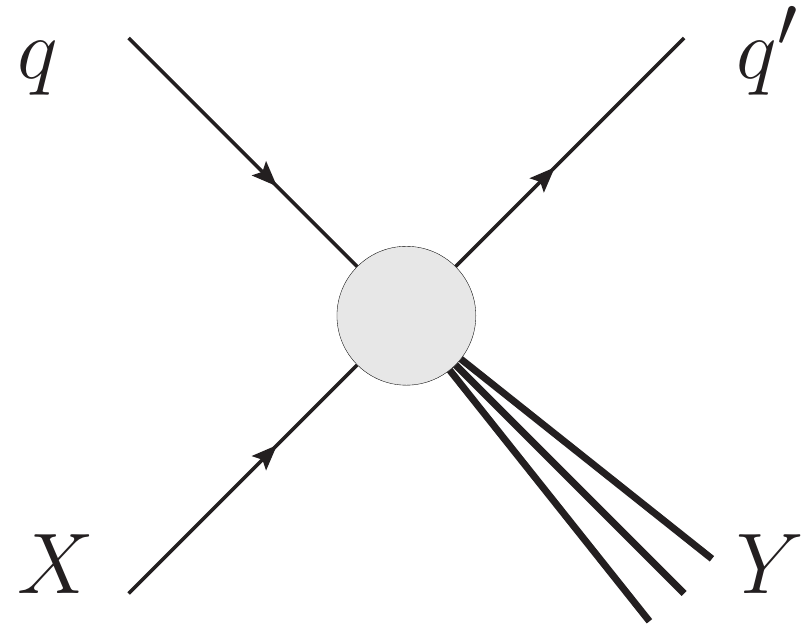}
\hspace{0.15\linewidth}
\begin{picture}(10,100)(0,0)
\put(20,75){\makebox[.2\linewidth]{$P_q=\left(E,\,\vec{0},\,E\right)\;\;\;P_X=\left(E_X,\,\vec{0},\,-E\right)$}}
\put(20,45){\makebox[.2\linewidth]{$P_{q'}=\left(\sqrt{(1-x)^2E^2+p_\bot^2},\,\vec{p_\bot},\,E(1-x)\right)$}}
\put(20,15){\makebox[.2\linewidth]{$\vec{p}_\bot=\{p_\bot\cos\phi,p_\bot\sin\phi\}$}}
\end{picture}
%\vspace{1cm}
\caption{Pictorial representation of the $qX\rightarrow q' Y$ process, 
which defines our parametrization of the quark momenta $P_q$ and 
$P_{q'}$ in the center of mass frame. The momentum fraction $x$ is taken 
to be in the $(0,1)$ interval and far enough from the extremes in order 
for the first condition in eq.~(\ref{limit}) to be satisfied.}\label{mom_par}
\end{figure}
%%%%%%%%%%%%%%%%%%%%%%%%%%%%%

Let us consider a generic scattering process, of the type ${q\, X\rightarrow q'\, Y}$, 
in which a massless quark (or a lepton) $q$ scatters on an unspecified initial particle 
$X$, gets converted into a second quark $q'$ possibly of different charge and 
produces an unspecified final state $Y$. The processes we are interested in are the 
``weak'' ones, in which the only relevant interaction of the quarks is provided by the 
standard minimal weak coupling with the $W$ bosons. To simplify the discussion we ignore the QCD interactions and set to zero the hypercharge coupling  $g'$. In practice, 
we assume that the only relevant quark vertex is given by
\beq
\displaystyle
{\mc L}_q\,=\, \frac{g}2\,\overline{q}_L\sigma^a\gamma^\mu q_L W_\mu^a \,.
\label{qint}
\eeq
We see no obstruction to include hypercharge vertices and we do not expect 
any qualitative change in the conclusions.

Apart from the minimality of the quark's interaction in eq.~(\ref{qint}), which however is very 
well motivated experimentally and therefore does not constitute a limitation \footnote{In the 
kinematic regime where EWA applies, the $Wq\bar q$ interaction is probed at a low virtuality where the possible effect of new physics, parametrized by higher dimension operators, is negligible.},
we will not need 
any other assumption on the theory that governs the dynamics of the weak bosons. Given that we aim, 
as explained in the Introduction, to use  EWA as a tool for probing the presently unknown EWSB 
sector, it is important 
to stress (and to prove, as we will do in section~3) that the validity of  EWA does not rely on any 
assumption on EWSB. The sector responsible for the $W$ interactions is generic and, given the typical 
energy $E$ of the scattering process, can be depicted as a set of particles with mass below or around $E$ that interact 
through a set of couplings $c_i$, possibly originating by integrating out states that are much heavier 
than $E$. For example, suppose that the EWSB sector just consists of the standard Higgs model. 
If the Higgs is light ($m_H\lesssim E$), the appropriate description of the $W$s and Higgs 
interactions is provided by the linear Higgs Lagrangian, and only contains renormalizable 
couplings. If on the contrary the Higgs is heavy ($m_H\gg E$), the $W$ interactions are dictated 
by the non-renormalizable $\sigma$-model obtained by integrating it out; our general analysis will 
encompass both possibilities. 

\begin{figure}
\centering
\includegraphics[width=0.46\linewidth]{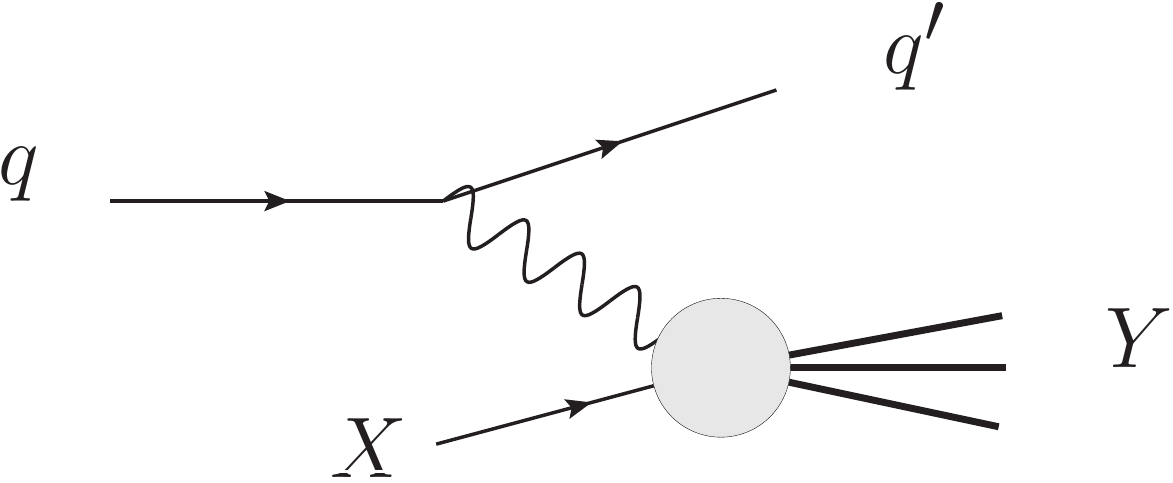}
\includegraphics[width=0.46\linewidth]{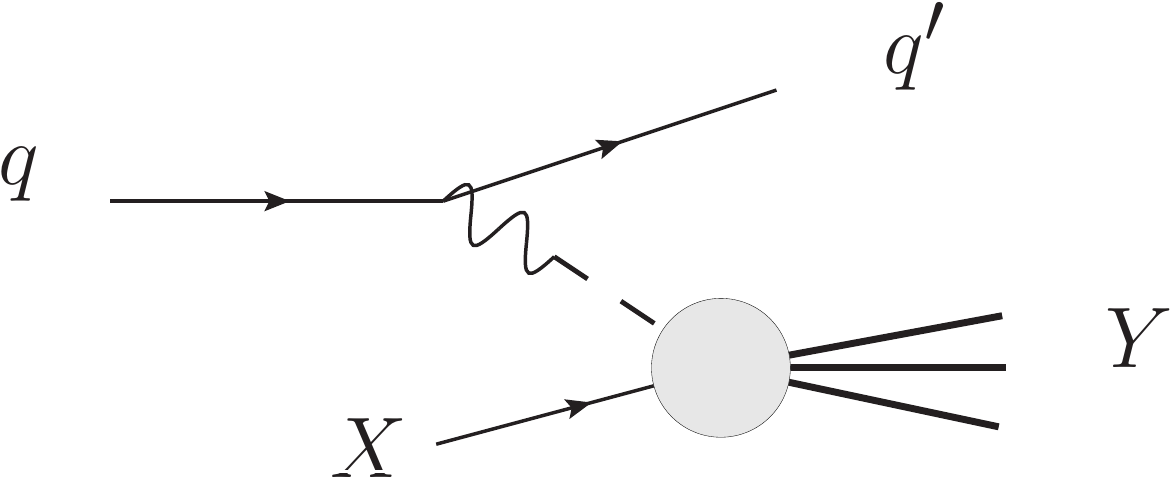}
\caption{The topology of the scattering contributions 
to the $qX\rightarrow q'Y$ process, the second diagram contains 
a mixed gauge-goldstone propagator which arises because we work in the axial 
gauge.}
\label{scatt}
\end{figure}

The essence of EWA is that the complete ${q\, X\rightarrow q'\, Y}$ process  is 
equivalent, at high enough energies and under particular kinematic conditions, to the $WX\rightarrow Y$ 
hard sub-process initiated by an on-shell $W$ boson, convoluted with the splitting function that describes 
the collinear emission $q\rightarrow q'W$.  EWA is a particular example of the general phenomenon of 
factorization which typically takes place in the limit of highly energetic and forward splitting. We 
therefore expect EWA to hold when the energy $E$ of the process is much larger than the mass 
$m$ of the $W$ and in the special kinematic 
regime in which the quark undergoes, in the center of mass of the collision, a small angular deflection while losing a  sizable fraction of its initial energy. 
More precisely, in the parametrization of the momenta defined by Fig.~\ref{mom_par}, the regime relevant for 
EWA is
\beq
\displaystyle
E\,\sim\,x E\sim (1-x)E\,,\;\;\;\;\;\;\;\;\;\;
\delta_m\equiv\,\frac{m}{E}\,\ll \,1 \,, \;\;\;\;\;\;\;\;\;\;\delta_\bot\equiv\,\frac{p_\bot}{E}\,\ll \,1 \,, 
\label{limit}
\eeq
where no assumption is made, a priori, on the relative size of $p_\bot$ and $m$. In that kinematic regime, 
let us consider, among the different diagrams contributing to  ${q\, X\rightarrow q'\, Y}$, 
the ``scattering'' diagrams depicted in Figure~\ref{scatt}. At the $W\bar q' q$ vertex, on-shell $q$ and $q'$ 
are respectively annihilated and created, together with the creation of  an off-shell virtual $W$ of momentum
\beq
\displaystyle
K\,\equiv\,P_q\,-\,P_{q'}\,=\,\left(\sqrt{
x^2E^2\,+\,p_\bot^2\,+\,m^2\,-\,V^2
},\,-\vec{p_\bot},\,x E\right)\,,
\label{vwmom}
\eeq
where the virtuality (off-shelness) $V$ is given by
\beq
\displaystyle
V^2\,=\,m^2\,-\,K^2\,\simeq\,m^2\,+\,\frac{{p_\bot}^2}{1-x}\left[1+\ord\left({\frac{p_\bot}{E}}\right)^2\right]
\,.
\label{virt}
\eeq
Once emitted, the virtual $W$ enters the  $WX\rightarrow Y$ sub-process, which we 
assume to be characterized by a hardness $H$ of order $E$. The 
hardness corresponds to the time scale of the $WX\rightarrow Y$ sub-process. The virtuality $V$, instead,  sets  
the time interval $\Delta T$ during which the virtual $W$ is indistinguishable, due the 
Uncertainty Principle, from an on-shell physical $W$. The correction to the $W$ energy due 
to off-shellness is indeed $\Delta E\sim V^2/E$, from which $\Delta T\sim E/V^2$. Given that 
 $V$ is much smaller than $E$ ($V^2/E^2={\mc O}(\delta_m^2+\delta_\bot^2)$) because of eq.~(\ref{limit}), 
the time  scale $1/E$ of the $WX\rightarrow Y$ sub-process  is much shorter than $\Delta T$ meaning that it 
must be possible to regard the $W$ as an on-shell particle during the hard scattering. The 
full process must factorize in terms 
of the on-shell $WX\rightarrow Y$ cross section.

Technically,  EWA can be stated as follows \cite{Dawson:1984gx}. Consider the exact 
cross-section \mbox{$d\sigma(qX\rightarrow q'Y)$}, integrated over the azimuthal 
 angle $\phi$ (see Fig.~\ref{mom_par}) of the final quark but completely differential 
in all the other kinematic variables. In the 
$\delta_{m,\,\bot}\hspace{-4pt}\rightarrow0$ limit, according to EWA the unpolarized cross-section
$d\sigma(qX\rightarrow q'Y)$ reduces to 
\begin{eqnarray}
\displaystyle
\frac{d\sigma_{\textrm{EWA}}(qX\rightarrow q'Y)}{dx dp_\bot}\,=\,
\frac{C^2}{2\pi^2}     && \left\{ 
      \,\,      f_{+}(x,p_{\bot})          \times     d\sigma(W^Q_+ X\rightarrow Y)+ \right.  \nonumber\\
&&  +    f_{-}(x,p_{\bot})           \times     d\sigma(W^Q_- X\rightarrow Y)+               \nonumber\\
&&  \left. +   f_{0}(x,p_{\bot})         \times     d\sigma(W^Q_0 X\rightarrow Y)  \right\}  \,,
\label{EWA}
\end{eqnarray}
where each line is associated with the exchange of an equivalent $W$ boson of a different polarization $r=\pm,0$. The three splitting functions $f_{\pm}$ and $f_0$ describe the collinear emission of the polarized equivalent $W$ and $d\sigma(W^Q_r X\rightarrow Y)$ denotes the totally differential polarized cross-section of the $W^Q_r X\rightarrow Y$ hard sub-process. In the sub-process the equivalent $W^Q_r$ is treated as a perfectly physical on-shell particle of four-momentum
\be
\displaystyle
K_W\,=\,\left(\sqrt{x^2E^2+m^2},\,\vec0,\,xE\right)\,.
\label{wmon}
\eeq
The variable $x$ thus corresponds to the fraction of longitudinal momentum (which approximately coincides with the energy for $E\gg m$) carried away by the equivalent $W$ from the fermionic line. Explicitly, the splitting functions are given by
\begin{eqnarray}
f_{+}(x,p_{\bot})&=&  {(1-x)^{2} \over x} {{p_{\bot}^{3}} \over (m^{2}(1-x)+p_{\bot}^{2})^{2}}  \, ,\nonumber\\
f_{-}(x,p_{\bot})&=&  { 1 \over x  } {{p_{\bot}^{3}} \over (m^{2}(1-x)+p_{\bot}^{2})^{2}} \, ,  \nonumber\\
f_{0}(x,p_{\bot})&=&  {  (1-x)^{2}\over x }  {{2 m^{2} p_{\bot}} \over (m^{2}(1-x)+p_{\bot}^{2})^{2}} \, .
\label{Ftildadef}
\end{eqnarray}
$Q=\pm1,0$ appearing in the previous equations   denotes the electric charge difference between $q$ and $q'$ and obviously corresponds to the charge of the equivalent $W$ emitted in the splitting. Depending on $Q$, the coefficient $C$ in eq.~(\ref{EWA}) reads
\begin{eqnarray}
&&C\,=\,\frac{g}4\;\;\;\;\;{\textrm{for}}\;\;\;\;\;Q=0\,,\nonumber\\ 
&&C\,=\,\frac{g}{2\sqrt{2}}\;\;\;\;\;{\textrm{for}}\;\;\;\;\;Q=\pm1\,.
\label{Ci}
\end{eqnarray}
For  processes initiated by two quarks, or by a quark and an antiquark $q_i {\overline{q}}_j\to q_i'{\overline{q}}_j' X $ which both collinearly produce two equivalent $W$'s, EWA obviously generalizes to
\begin{eqnarray}
\frac{ d\sigma_{EWA} \left(  q_{i} {\overline{q}}_j  \to q_{i}' {\overline{q}}_j' X \right)}{dx_{i}dx_{j}dp_{\bot,i}dp_{\bot,j}  } = \sum_{r,s}  && { C_{i} ^{2} \over  2 \pi^{2}} { C_{j} ^{2} \over  2 \pi^{2}} \,  f_{r}(x_{i},p_{\bot,i}) f_{-s}(x_{j},p_{\bot,j})  \times d\sigma \left(   W_{r}^{Q_{i}}W_{s}^{Q_{j}}\to  X   \right),
\label{EWA2to4}
\end{eqnarray}
where the inversion of the equivalent $W$ polarization in the anti-quark splitting function follows from  $CP$ conjugation.

The derivation of eq.~(\ref{EWA}), which we will describe in section~3, basically consists of a Laurent series expansion of the exact 
scattering amplitude in the parameters $\delta_\bot=p_\bot/E$ and $\delta_m=m/E$, a procedure that will also allow us to establish the scaling of the 
corrections to the approximation. We will find that the relative corrections are typically quadratic in $\delta_{m}$ and in $\delta_{\bot}$, {\it{i.e.}} 
\begin{equation}
\delta_{EWA}\,=\,\textrm{Max}\left[\delta_m^2,\,\delta_\bot^2\right]\,,
\label{EWAcorr}
\end{equation}
even though they can be enhanced in some particular condition as we will discuss in section~\ref{correwa}. For instance, in the very low-$p_\bot$ region $p_\bot\ll m$ the error scales in some cases as \mbox{$\delta_{EWA}=(m^2/p_\bot E)^2$} and in particular for $p_\bot\simeq m^2/E$  EWA fails even if $\delta_m$ and $\delta_\bot$ are extremely small. We will confirm the above result in the following section, by computing the corrections explicitly in the case of the $WW\rightarrow WW$ hard sub-process.

Before moving forward, two important comments are in order. The first is 
that for the intuitive derivation of the EWA previously discussed it has been 
crucial to assume that the $WX\rightarrow Y$ sub-process is genuinely 
hard, with a hardness $H\sim E$ and that it contains no other 
softer scales. 
This excludes, for instance, the case where  the momenta of the final particles 
$Y$ become soft, or collinear among each other, 
with $P_X$ or with the virtual $W$ momentum $K$. In order to apply 
EWA all those kinematic configurations must be avoided by suitable 
cuts on the $Y$ momenta. In the processes considered in the 
following section, for instance, the region of forward $WW\rightarrow WW$
scattering is soft because it is characterized by a low value of the $t$ 
variable and will have to be excluded by a hard cut on the 
transverse momentum $P_\bot^W$ of the final $W$'s. 
Moreover, given that factorization relies on the hierarchy among 
the virtuality of the soft splitting and the 
hardness $H$ of the sub-process, one might expect that a numerically more precise estimate of the corrections to the EWA could be obtained 
by comparing $p_\bot$ and $m$ with $H$ instead of $E$, {\it{i.e.}} by 
redefining $\delta_m=m/H$ and $\delta_\bot=p_\bot/H$. We will verify this 
expectation in the following section, where $H=P_\bot^W$.

The second comment is that the intuitive argument in favor of the EWA implicitly 
relied on a gauge choice because of its starting point, which 
consisted in interpreting the $W$ propagator as the exchange of 
off-shell but otherwise perfectly physical $W$ quanta. This 
interpretation is only valid in ``physical'' gauges while in 
a generic one, including for instance the covariant gauges, 
extra unphysical states propagate and the scattering diagrams 
are not endowed with the physically transparent interpretation 
outlined above. This suggests that the task of providing a 
technical proof of the EWA might be easier to accomplish if working 
in a physical gauge, and this is indeed what we will do 
in section~3, where we will choose the axial gauge. 

\subsection{Numerical Comparison}
\label{nuns}

In order to get a first confirmation of the validity of the EWA approximation in eqs.~(\ref{EWA},\ref{EWA2to4}),  we will now perform a numerical comparison of $d\sigma_{EWA}$ with the exact differential cross section
$ d\sigma_{EXACT}$ in some explicit examples. We will quantify the accuracy of the approximation by computing
\begin{equation}
\delta_{EWA}\equiv 2 \, \frac{d\sigma_{EXACT}-d\sigma_{EWA}}{d\sigma_{EXACT}+d\sigma_{EWA}}\,,
\label{eq:errore}
\end{equation}
at some reference points in  phase space.
 We will consider two processes: the $2\to3$ 
\begin{equation}
uW^{-}\to dW^{+}W^{-}\,.\label{eq:2to3}
\end{equation}
and the $2\to4$ quark/antiquark scattering
\begin{equation}
u\bar{c}\to d\bar{s}W^{+}W^{-}\,,\label{eq:2in4}
\end{equation}
where we have chosen initial quarks of different families (with vanishing Cabibbo angle) just for simplicity, in order to avoid a proliferation of diagrams with $q\,{\overline{q}}$ annihilation. 

Let us describe the kinematics for the approximated and for the   exact cross section.
The hard $2\rightarrow2$ cross section which appears in  EWA, $d\sigma \left(   W_{r}^{+}W_{s}^{-}\to W^+ W^-   \right)$, is computed by taking as independent kinematic variables the polar angle $\theta$ in the center  of mass (C.M.)  of the collision and the azimuthal angle $\phi_W$ of the outgoing $W^{+}$. The C.M. energy of the $W^+W^-$ system is instead fixed by the momenta of the incoming equivalent $W$'s which, in analogy with  eq.(\ref{wmon}), read
\begin{eqnarray}
&&\textrm{for the }2\to3\textrm{:}\;\;\; K_{+}=(\sqrt{m^{2}+E^{2}x^{2}},\,0,\,0,\, Ex)\,,\;\;\; K_{-}=(\sqrt{m^{2}+E^{2}},\,0,\,0,\, -E )\,,\nonumber\\
&&\textrm{for the }2\to4\textrm{:}\;\;\; K_{+}=(\sqrt{m^{2}+E^{2}x^{2}},\,0,\,0,\, Ex)\,,\;\;\; K_{-}=(\sqrt{m^{2}+E^{2}y^{2}},\,0,\,0,\, -Ey )\,,
\label{EWAM}
\end{eqnarray}
where $y$ is the energy fraction of the equivalent $W^-$ coming from the anti-quark splitting. The remaining relevant kinematic variables characterizing the $2\to 3 $ and $2\to 4$ processes (see eqs ~(\ref{EWA}) and  (\ref{EWA2to4})) are just the absolute values of  the outgoing quarks transverse momenta. These  
are the same as for the kinematics of the exact process that we describe below. In EWA one integrates on the quarks' azimuthal angles, and thus  the result does not depend  on $\phi_W$.

The kinematic of the exact $2\rightarrow4$ process  is  parametrized as follows. The incoming quark and anti-quark momenta are
\begin{eqnarray}
P_{u} & = & \left(E,0,0,E\right)\, ,  \label{puc}\\
P_{\bar c} & = & \left(E,0,0,-E\right)\,, 
\end{eqnarray}
while for the outgoing ones we have, in analogy with Fig.~\ref{mom_par}, 
\begin{eqnarray}
P_{d} & = & \left(\sqrt{E^{2}(1-x)^{2}+p_{\bot}^{2}},p_{\bot}\sin(\phi),p_{\bot}\cos(\phi),E(1-x)\right)\,, \label{pdownkinEWA24}   \\
P_{\bar s} & = & \left(\sqrt{E^{2}(1-y)^{2}+q_{\bot}^{2}},q_{\bot}\sin(\psi),q_{\bot}\cos(\psi),-E(1-y)\right)\,.
\end{eqnarray}
In order to compare with EWA we must however integrate over the azimuthal angles  of the outgoing quarks
$\phi$ and $\psi$.  As for the outgoing $W^{+}W^{-}$ system, we describe it, as before, by the  angles $\theta$ and $\phi_W$ of the $W^+$ in the $W$-pair center of mass frame.  Notice that now, unlike for the $2\to2$ EWA case, the boost of the $W^{+}W^{-}$ system is not directed along the $z$-axis but has a transverse component recoiling against the quark  transverse momenta $p_\bot$ and $q_\bot$. Nevertheless, by azimuthal symmetry, after  integrating over $\phi$ and $\psi$, the remaining dependence on $\phi_W$ is trivial.

As for the exact $2\to 3$ process, we describe the quark kinematics by the same $P_u$ and $P_d$ of eqs.~(\ref{puc},\ref{pdownkinEWA24}). In strict analogy we then define the $WW$ system variables and integrate over the $d$-quark azimuthal angle $\phi$. In the end,
for the $2\to 3$ process the independent kinematic variables we fix in order  to compute $\delta_{EWA}$ are  $p_\bot, x$ and $\theta$. For the $2\to 4$ process there are two additional variables $y, q_\bot$ associated with the $\bar c\to \bar s$ quark line. The dependence on $\phi_W$ is trivial.
It should be noticed that the comparison between  EWA and the exact cross section is \emph{not} performed at exactly coincident kinematic points. This is because   in the  kinematics of EWA   the conservation of both transverse momentum and energy is  violated by the quark splitting into the equivalent vector bosons (eq.~(\ref{EWAM})).  So at corresponding values of 
 $\theta$ and $\phi_W$ the outgoing  $W$ $4$-momenta differ in EWA and in the exact kinematics. The differences are however of order $p_\bot^2/E^2$, $m^2/E^2$. That is precisely the order at which we expect the approximation to hold.
  
Two final  comments on the analysis that follows are in order. We shall also use a variable $k_T$,
 function of the independent kinematic variables, which corresponds to the transverse momentum of the 
outgoing $W^+$ in EWA kinematics. By defing $s_{WW}= (K_-+K_+)^2$ the CM square energy in the EWA kinematics we have
\begin{equation}
 k_{T} = \sin\theta \, {\sqrt{s_{WW} - 4m^{2} \over 4}}\,, \label{kTthetastar}\,.
 \end{equation}
Of course this quantity, as a function of $\theta, x, p_\bot$ ($y,q_\bot$) differs from the outgoing transverse momenta of either $W$'s in the exact kinematics by an amount $O(\delta_\bot^2,\delta_m^2)$. We will call this variable the pseudo-transverse-momentum. In all the analysis we shall pick a reference point where
 the Higgs boson $H$ has a  mass $m_{H}=2m$ and consider vanishing hypercharge, compatibly with eq.~(\ref{qint}). The scattering amplitudes are computed with the FeynArts/FormCalc system \cite{FeynArts} and the numerical integration over the jets' azimuthal angle is performed using the \textsc{Cuba} library \cite{Cuba}.
 
%%%%%%%%%%%%%%%%%%%%%%%%%%%%
\subsubsection{Corrections as a function of the hardness of the final state $W$}
%%%%%%%%%%%%%%%%%%%%%%%%%%%%

\begin{figure}
\begin{centering}
\includegraphics[width=0.66\linewidth]{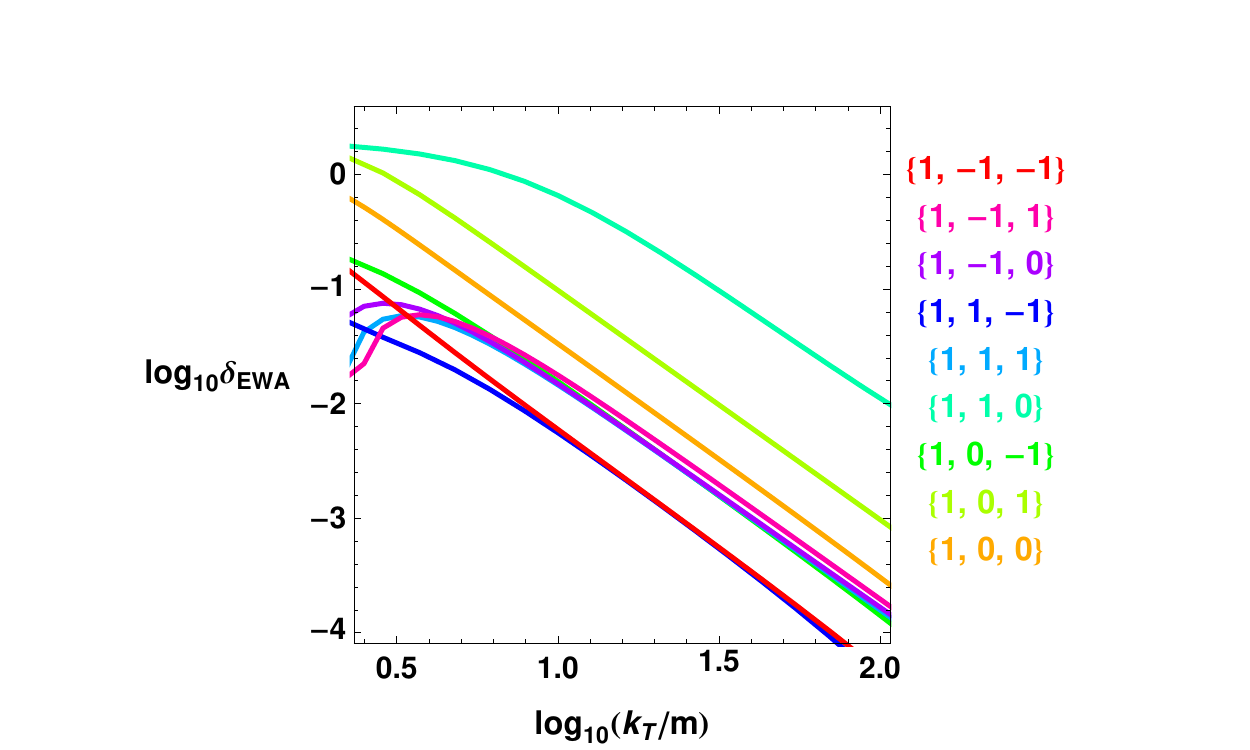}
\par\end{centering}
\begin{centering}
\includegraphics[width=0.66\linewidth]{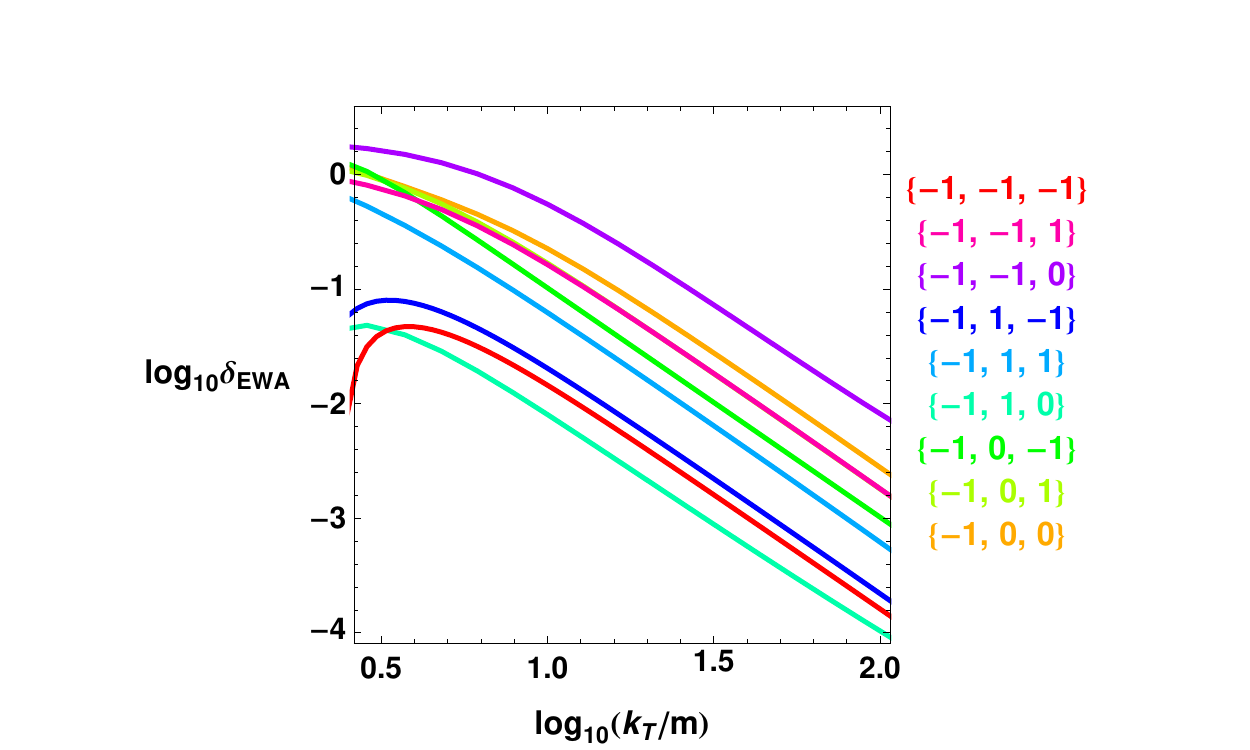}
\par\end{centering}
\begin{centering}
\includegraphics[width=0.66\linewidth]{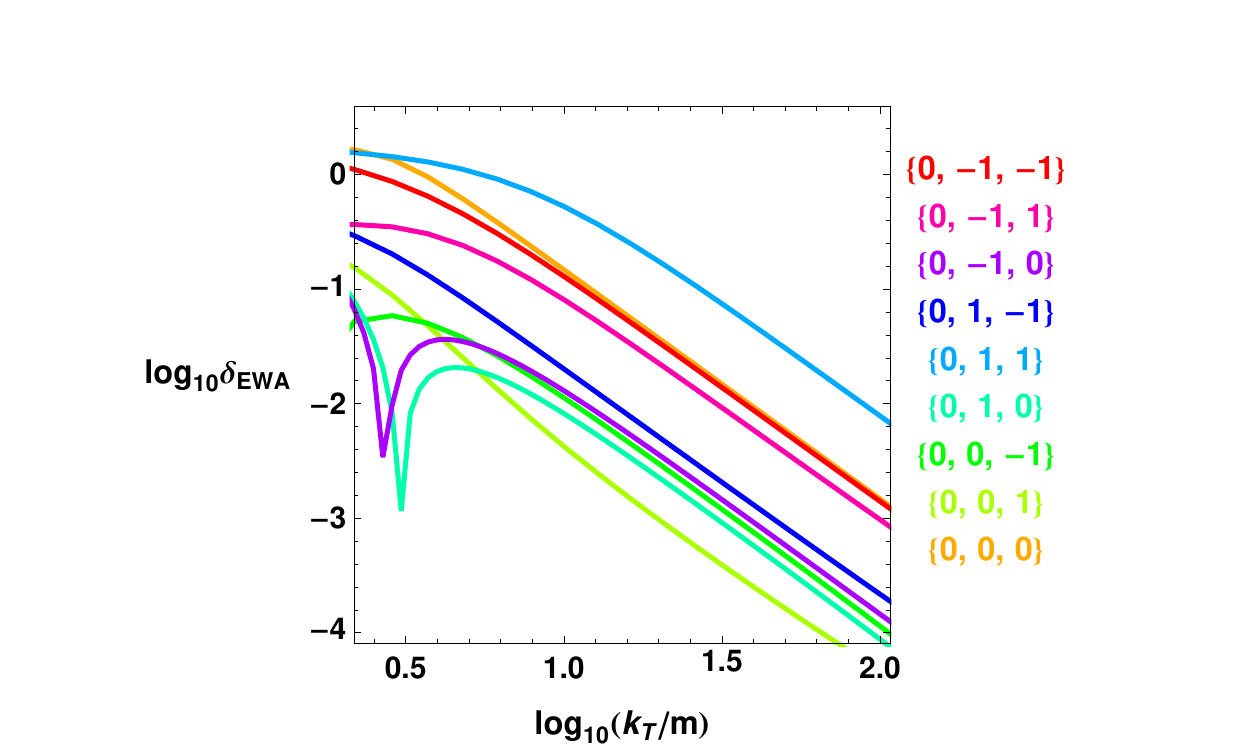}
\par
\end{centering}
\caption{\label{fig:delta2in3p}Accuracy of  EWA for the $2\to3$ process in eq.~(\ref{eq:2to3}) as a function of $k_{T}/m$ for fixed kinematics given by eq.~(\ref{eq:fixedKinematics2in3}). The helicities of the external $W$ bosons in each process are $\{ \lambda(W^{-}_{in}),\, \lambda(W^{+}_{out}),\,\lambda(W^{-}_{out}) \}$ as  indicated by the colored labels at the right of each plot.}
\end{figure}

As explained in Section \ref{BasicPicture}, a necessary condition for  factorization  is a hierarchy between the virtuality (or the hardness) of the $q\to q'W'$ splitting and that of the $WW\to WW$ subprocess. As such we compute $\delta_{EWA}$  as a function of  the pseudo-tranverse-momentum $k_{T}$, by keeping  the C.M. scattering angle $\theta$ and the outgoing jets transverse momenta fixed  while increasing the center of mass energy. In particular  for the process in eq.(\ref{eq:2to3}) we show our computation of $\delta_{EWA}$
as a function of $k_{T}/m$ in Figure \ref{fig:delta2in3p} for a fixed kinematics given by 
\begin{equation}
p_\bot=45\textrm{ GeV,\,}\;\;x=0.65,\,\;\; \sin\theta=0.95
%,\,\phi_{W}=\frac{\pi}{2}
\,.\label{eq:fixedKinematics2in3}
\end{equation}
The three panels show $\delta_{EWA}$ for all the 27 combinations
of helicities of the external states of the process. In all cases
 EWA captures the leading behavior of the exact cross-section. From the figure one can read that $\delta_{EWA}$
scales like $1/k_{T}^{2}$, in agreement with the expectations. Notice however that the corrections are numerically larger than the estimate $\delta_{EWA}\simeq m^{2}/k_{T}^{2}$ (notice that $ p_{\bot}<m$, so that the corrections are dominated by $m$) in eq.~(\ref{EWAcorr}). This signals the presence of a large numerical coefficient, typically of the order of a few. It would be interesting to investigate the impact of this numerical enhancement in an experimentally realistic situation like the LHC at 14 TeV.

%%%%%%% FIGURE %%%%%%
%%%%%%%%%%%%%%%%%%%%%%%%%%%%%%%%%%%%

\begin{figure}
\begin{centering}
\includegraphics[width=0.66\linewidth]{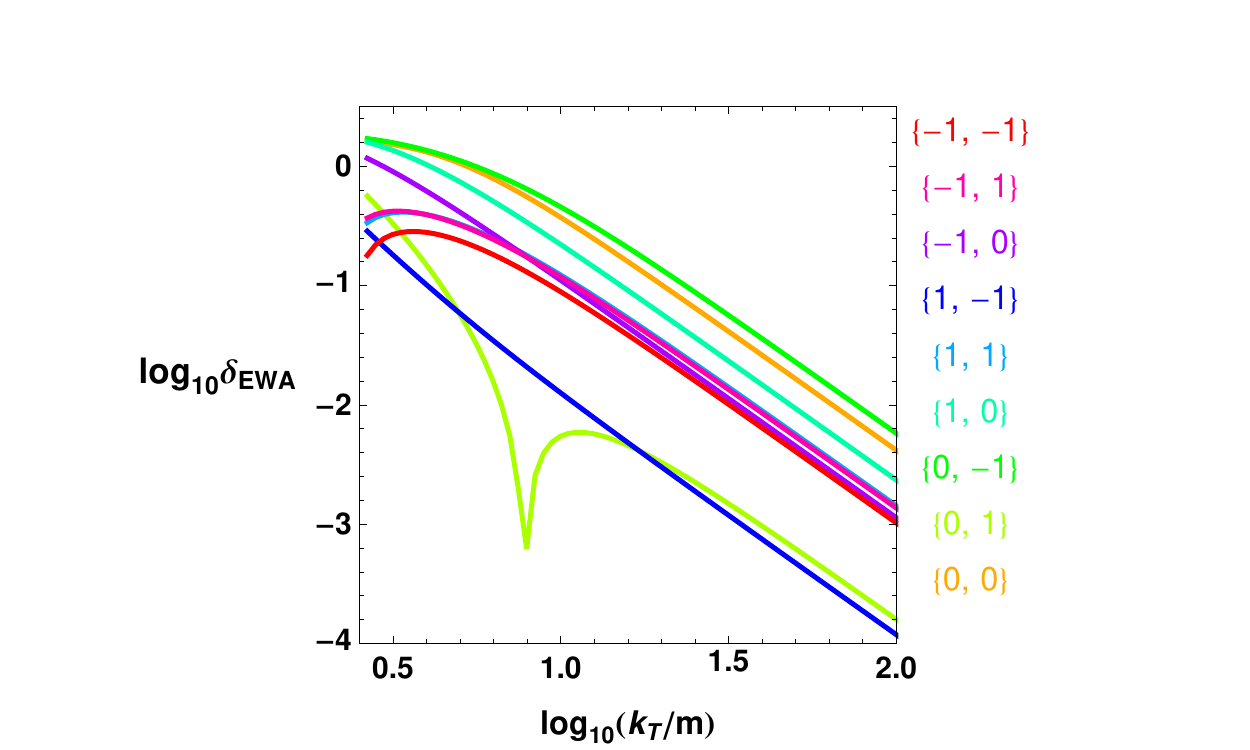}
\par\end{centering}
\caption{\label{fig:delta2in4p}Accuracy of  EWA for the $2\to4$ process eq. (\ref{eq:2in4}) as a function of $k_{T}/m$
for fixed kinematics given by eq.(\ref{eq:fixedKinematics2in4}).  The helicities of the external $W$ bosons in each process are $\{ \lambda(W^{+}_{out}),\,\lambda(W^{-}_{out}) \}$ as  indicated by the colored labels at the right of the plot. }
\end{figure}

The analogous computation of $\delta_{EWA}$ for the process in eq.(\ref{eq:2in4}) is shown in Figure~\ref{fig:delta2in4p}
for a fixed kinematics given by 
\begin{equation}
p_{\bot}=20 \textrm{ GeV,\,}\;\;q_{\bot}=10 \textrm{ GeV,\,}\,\;\;x=0.4,\,\;\;y=0.6,\, \;\;\sin\theta={\sqrt{3} \over 2}
%,\,\phi_{W}=1.1
\,.\label{eq:fixedKinematics2in4}
\end{equation}
Also in this case the scaling of the corrections to the EWA is such that $\delta_{EWA}\sim k_{T}^{-2}$, in agreement with eq.~(\ref{EWAcorr}).

\begin{figure}
\begin{centering}
\includegraphics[width=0.66\linewidth]{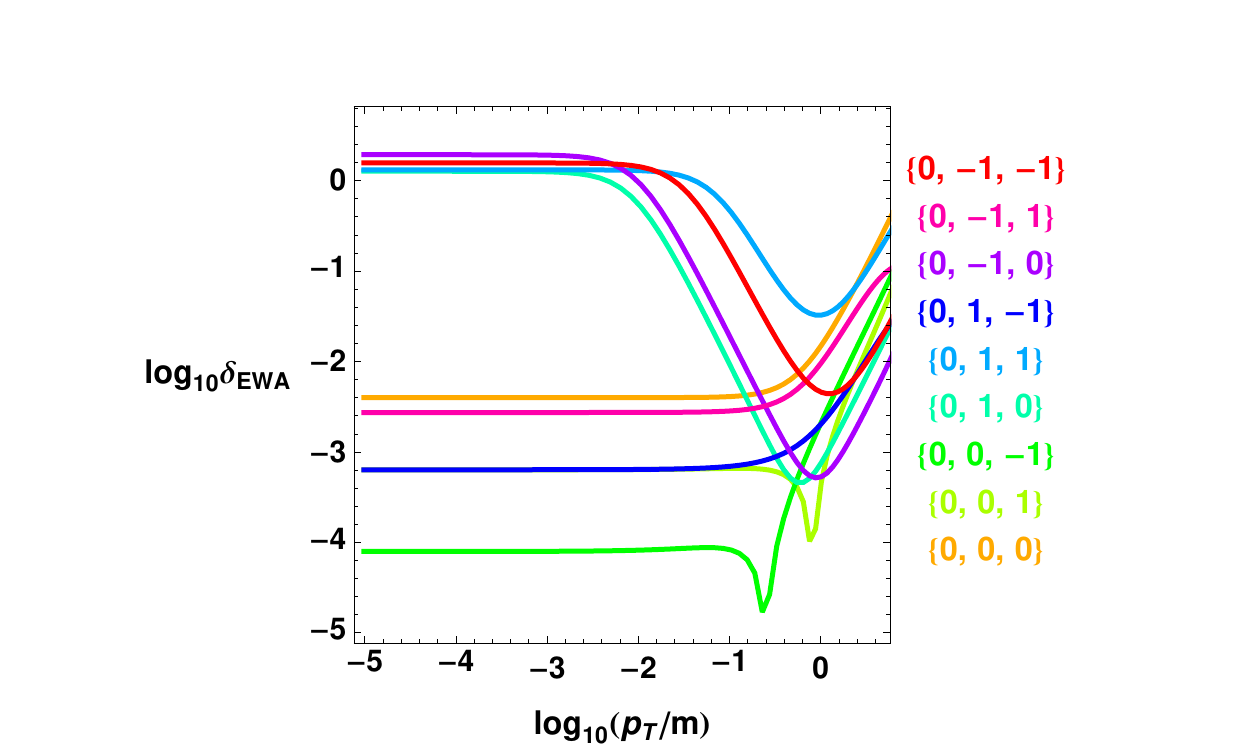}
\par\end{centering}

\begin{centering}
\includegraphics[width=0.66\linewidth]{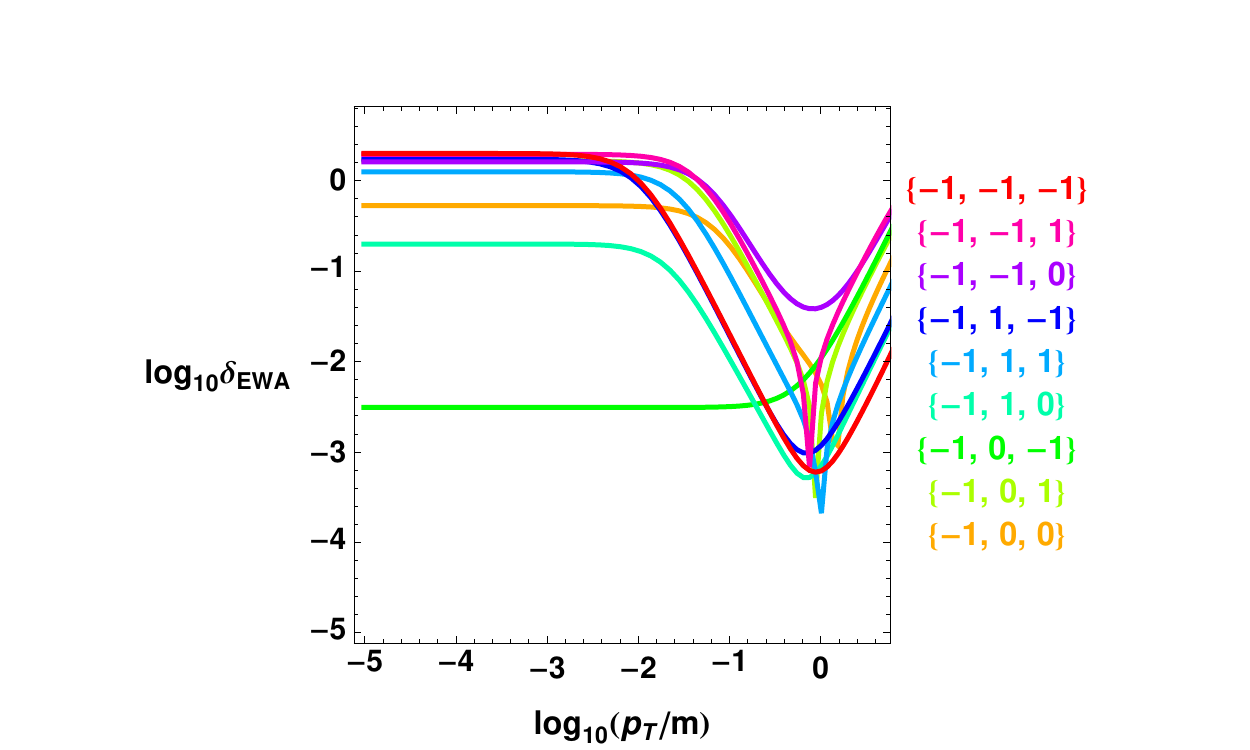}
\par\end{centering}

\begin{centering}
\includegraphics[width=0.66\linewidth]{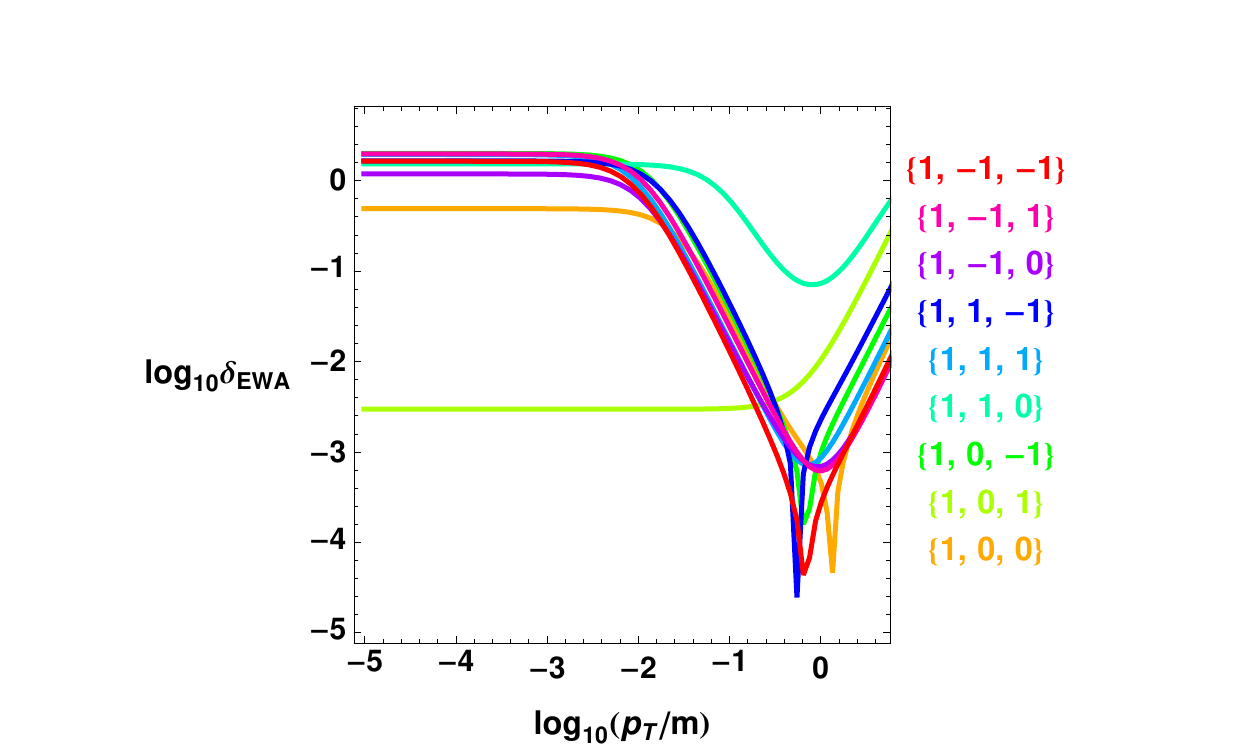}
\par\end{centering}

\caption{\label{fig:PTdelta2in3}Accuracy of  EWA as a function of $p_{T}/m$ for fixed kinematics as in eq.(\ref{eq:kin2in3pTvariation}). The helicities of the external $W$ bosons in each process are $\{ \lambda(W^{-}_{in}),\, \lambda(W^{+}_{out}),\,\lambda(W^{-}_{out}) \}$ as  indicated by the colored labels at the right of each plot.}
\end{figure}

\begin{figure}
\begin{centering}
\includegraphics[width=0.66\linewidth]{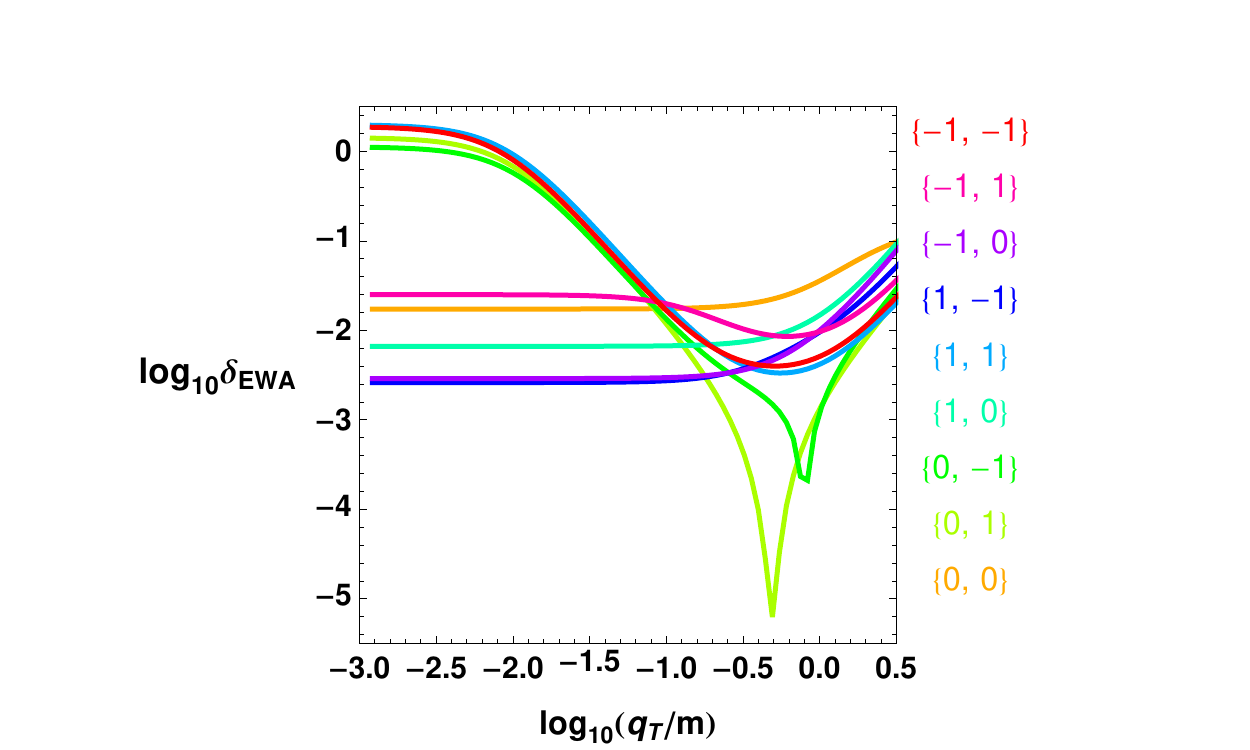}
\par\end{centering}
\caption{\label{fig:PTdelta2in4}Accuracy of  EWA as a function of $q_{\bot}/m$
for fixed kinematics given in eq.(\ref{eq:kin2in4qTvariation}).  The helicities of the external $W$ bosons in each process are $\{ \lambda(W^{+}_{out}),\,\lambda(W^{-}_{out}) \}$ as  indicated by the colored labels at the right of the plot.}
\end{figure}

%%%%%%%%%%%%%%%%%%%%%%%%%%
\subsubsection{Corrections as a function of the hardness of the jets}
%%%%%%%%%%%%%%%%%%%%%%%%%%
The results shown so far are relative to points of phase-space where, according to eq.~(\ref{EWAcorr}), the corrections to EWA are dominated by $m$, as the chosen transverse momenta of the jets in eqs.~(\ref{eq:fixedKinematics2in3}) and (\ref{eq:fixedKinematics2in4}) are relatively low. In order to fully check the structure of the corrections in eq.~(\ref{EWAcorr}) we also have to study  the dependence of 
$\delta_{EWA}$ on the quark transverse momenta. For the $2\to 3$ process  we perform this study by fixing
\begin{equation}
\label{eq:kin2in3pTvariation}
E=10 \textrm{ TeV},\;\; x=0.65,\,\;\; \sin\theta=0.86
%,\,\phi_{W}={\pi \over 2}
\,,
\end{equation}
and by computing $\delta_{EWA}$   as a function of  $p_{\bot}/m$, the results are reported in Fig.~\ref{fig:PTdelta2in3}. We notice, first of all, that $\delta_{EWA}$ grows quadratically with $p_\bot$ in the region $p_\bot>m$. In that region, as predicted by eq.~(\ref{EWAcorr}), the corrections must indeed  go like $p_{\bot}^2/k_T^2$. At low $p_\bot\ll m$ different behaviors are observed for different helicities of the external $W$'s. In some cases $\delta_{EWA}$ stabilizes to a constant value $\simeq m^2/k_T^2$. In some others $\delta_{EWA}$ scales like $1/p_\bot^2$ and for $p_\bot\simeq m^2/k_T$ EWA fails completely. The corrections can be estimated in this case as $\delta_{EWA}\simeq (m^2/p_\bot E)^2$ as anticipated in the discussion below eq.~(\ref{EWAcorr}). We will explain this behavior in great detail in section~\ref{correwa} and \ref{wwsec}. For the moment we anticipate that the only helicity channels for which $\delta_{EWA}$ is enhanced are those dominated by the exchange of a transverse equivalent $W$. The enhancement of the relative corrections at low $p_\bot$ follows from the $p$-wave suppression of the leading order which affects the transverse splitting functions in eq.~(\ref{Ftildadef}). For the $2\to 4$ process we perform the same check by fixing
\begin{equation}
\label{eq:kin2in4qTvariation}E=10 \textrm{ TeV},\,\;\;p_{\bot}=20 \textrm{ GeV,\,}\,\;\;x=0.4,\,\;\;y=0.6,\, \;\;\sin\theta=0.8
%,\,\phi_{W}=1.3
\,,
\end{equation}
and compute  $\delta_{EWA}$ as a function of $q_{\bot}/m$. The result is shown in Fig.~\ref{fig:PTdelta2in4} where we observe the same behavior as in the case $2\to 3$.

\section{The Equivalent ${\mathbf{W}}$ Boson \label{ewaproof}}

The results of the previous section provide an extremely non-trivial 
confirmation of EWA, but of course their validity is limited to a 
specific process, $WW$ scattering, in a specific model of EWSB, the 
Higgs model. The aim of the present section is to overcome these 
limitations by deriving the EWA formula analytically and showing that 
its validity is neither restricted to a specific class of scattering 
processes nor to a specific model of EWSB.

Before proceeding it is worth recalling what is the main methodological obstacle in establishing EWA, the one that often causes confusion.
When working with Feynman diagrams, the derivation of EWA, or of any factorization based on virtual quanta, amounts to trashing some diagrams and
approximating others. However individual Feynman diagrams do not have absolute physical meaning, as their functional form and their numerical value  depend on the chosen parametrization of the fields and, in particular, on the choice of gauge. In an arbitrary gauge,  each  trashed diagram, or the neglected off-shell piece in the diagrams of \mbox{Fig.~1a}, can  be arbitrarily sizeable, making the validity of the approximation not manifest at all. Of course the existence of a gauge where all the neglected terms are one by one numerically negligible would be sufficient to establish the approximation in all other gauges as well. To better appreciate how EWA stands as regards the above issue,  it is sufficient to focus on the diagram in Fig.~1a. In an obvious notation it can be written as (see also the discussion in section \ref{deriv})
\begin{equation}
\left \{J^1_\mu(V_1^2)\frac{P^{\mu\nu}(\vec p_1, V_1^2)}{p_1^2-m^2}\right \}\left \{J^2_\rho(V_2^2)\frac{P^{\rho\sigma}(\vec p_2, V_2^2)}{p_2^2-m^2}\right \}{\cal A}_{\nu\sigma}(V_1^2,V_2^2)\,,\label{poles}
\end{equation}
where we have highlighted the dependence of the amplitude for the $WW\to WW$ subprocess on the virtuality $V_{i}^2\equiv m^2-p_i^2$ of the intermediate $W$'s.
The above contribution to the total amplitude has a (pole) singularity at $V_i=0$. More importantly, it  is the only contribution featuring such a singularity. It therefore follows that the leading
behavior at the singularity, the residue, is fully accounted by the above term, and as such is gauge independent. To obtain the leading behavior we simply have to replace $V_1=V_2=0$ everywhere in the numerator in eq.(\ref{poles}). Notice that in order to reach the pole we formally need to go to complex momenta, but that is not  a problem. The residue will itself be a gauge invariant function of momenta. Its continuation back to real momenta is precisely the EWA amplitude. The $1/V_i$ poles in the EWA amplitude signify that the process of emission of the forward fermions and of   the virtual $W$ takes place over a length and time scales much bigger than those associated with the $WW\to WW$ subprocess. It is thus intuitive that the probability for the two phenomena should factorize. On the other hand, the other diagrams and the subleading terms in eq.(\ref{poles}) do not feature such a singularity and can be interpreted as transitions where all quanta (both the fermions and the W) are emitted in a single hard collision associated to a time and lenght scale   of order $1/E$. The presence of a single scale $E$ implies that all the quanta are emitted in low partial waves and smoothly populate   phase space: that smooth topology leads to negligible interference with the pronged topology that dominates the factorized amplitude.
 Notice, for instance, that when considering the Taylor expansion of the subamplitude
\beq
{\cal A}_{\rho\sigma}(V_1^2,V_2^2)={\cal A}_{\rho\sigma}(0,0)+{\cal A}_{\rho\sigma}^{(1,0)}(0,0) V_1^2+{\cal A}_{\rho\sigma}^{(0,1)}(0,0) V_2^2+\dots
\label{expand0}
\eeq
the terms beyond the first, which are neglected in EWA, combine with the propagator to give non-singular contributions. These have obvioulsy the same  structure as the contributions from all the other Feynman diagrams, like Fig.~1b. Thus  they cannot be independently gauge invariant. Indeed it was noticed long ago that, in both the unitary and $R_\xi$ gauges, the terms beyond the first in eq.(\ref{expand0}) display an unphysical growth with powers of $E/m$. The problem of these gauges is that, for different reasons, they feature singularities at $m\to 0$. The unitary gauge is singular in both the propagator and the external polarization vectors. The $R_\xi$ gauge has a well behaved propagator, however there is freedom in the choice of the polarization vectors decribing the longitudinal polarizations. The usual choice, where the Goldstones fields are set to zero, is also singular. Because of these singularities, it turns out that the formally subleading terms in eq.~(\ref{expand0}) are instead parametrically enhanced. In some cases there is one power  of $E/m$  for each external line in the $WW\to WW$ subprocess and one finds \cite{Kleiss:1986xp}
\beq
\frac{{\cal A}^{(1,0)}(0,0)}{{\cal A}(0,0)}\sim\frac{1}{E^2} \left (\frac{E}{m}\right )^4
\eeq
implying that at $E\gg m$, the numerical value of the diagram is completely dominated by the ``unphysical" gauge depend contribution. In the gauges where that happens power counting is not manifest, and in order to establish EWA one cannot just neglect some diagrams and approximate others. In those gauges the computation of the full set of diagrams is needed, in order to take into account for  numerically formidable, but conceptually trivial, cancellations.

 On the other hand the axial gauge, which was already adopted by KS, does not suffer from the above problems in that it is both physical and, above all, non-singular at $m\to 0$. The axial gauge is therefore the obvious choice to properly power-count all  diagrams and establish the validity of EWA at the quantitative level. It is the choice we shall make in this section, by focussing on the general process $q X\to q' Y$.   In subsection~3.1 we shall describe main technical tools, while in subsections 3.2 and 3.3 we shall expand the subamplitudes to establish EWA. In sections 3.4 and 3.5 we shall discuss the parametric dependence of the corrections to EWA.

\subsection{Axial Gauge and Feynman diagrams}
\label{sca}

As just explained, we shall work in the axial gauge. Following \cite{Kunszt:1987tk} and references therein, we fix the gauge
by picking a space-like 4-vector $n^\mu$  aligned along the beam direction
\beq
n^\mu\,=\,\left(0,\,0,\,0,\,1\right)^\mu\,.
\eeq
For each of the three $Q=\pm1,0$ charge eigenstates, the gauge-fixing condition 
\beq
n^\mu W_\mu^Q =0\,
\label{axcond}
\eeq
is strictly enforced  by a delta-function in the 
functional integral. That ensures  the field independence of the gauge variation of the gauge-fixing functional and the decoupling of the 
 ghosts. Unlike in the covariant $R_\xi$ 
gauges, in the axial gauge the mixings among gauge fields and Goldstones are not cancelled  and  
the propagators are thus non-diagonal. They can be regarded as matrices in the $5$-dimensional 
space formed by the $4$ $W_\mu$ components plus the Goldstone and are given 
by
\beq
\displaystyle
{\mc P}_{IJ}\,=\,\frac{i}{q^2-m^2}{\mc N}_{IJ}\,,
\eeq
with
\bea
\displaystyle
&&{\mc N}_{\mu\nu}(q)\,=\,-\,\eta_{\mu\nu}\,+\,\frac{q_\mu n_\nu+q_\nu n_\mu}{q_L}
\,+\,\frac{q_\mu q_\nu}{q_L^2}\,,\nn\\
&&
{\mc N}_{\mu g}(q)\,=\,{\mc N}_{g \mu}(q)^*\,=\,-i\,m\,\frac{1}{q_L}\left(n_\mu\,+\,\frac{q_\mu}{q_L}\right)\,,\nn\\
&&{\mc N}_{g g}(q)\,=\,1\,+\,\frac{m^2}{q_L^2}\,.
\label{nprp}
\eea
The three lines of the above equations describe respectively gauge-gauge, gauge-Goldstone 
and Goldstone-Goldstone propagation, while $q_L\equiv q\cdot n$ denotes the projection of $q$ 
along the beam direction. 
Obviously, because of the gauge-fixing (\ref{axcond}), the propagator matrix 
annihilates the $5$-vector $(n^\mu,\,0)$ and therefore has rank $4$. Moreover, as it will become apparent from the following discussion, for on-shell momentum, $q^2=m^2$, 
 the rank of the numerator ${\mc N}$ is further reduced down to 
$3$, corresponding to the physical polarizations of a massive vector. That is what makes the axial gauge 
``physical'' and therefore best suited for our proof of EWA.

\begin{figure}
\centering
\includegraphics[width=0.86\linewidth]{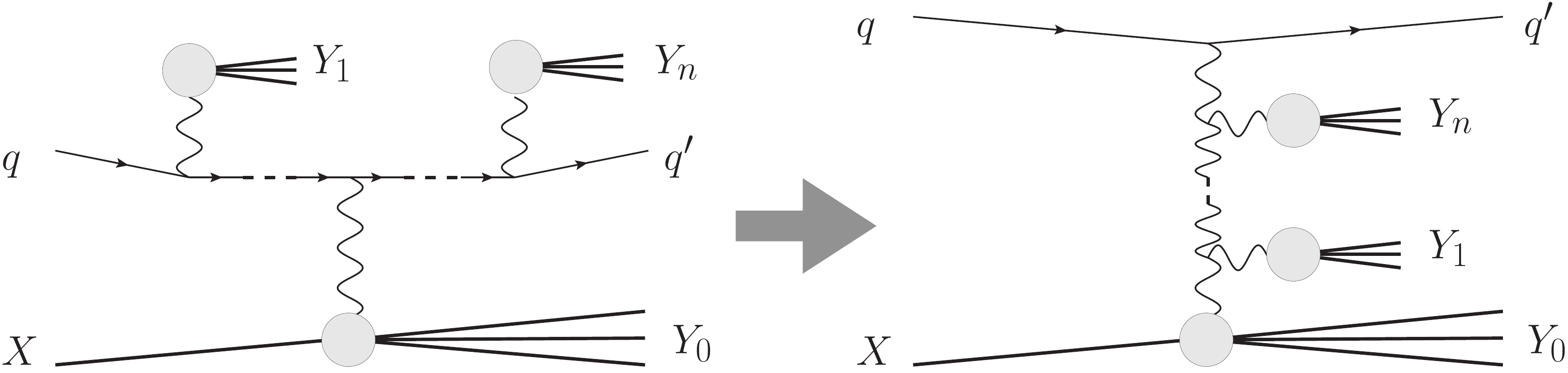}
\caption{(a) a representative radiation diagram: unlike the scattering diagrams, these diagrams do not have the $p_\bot/V^2$ or $m/V^2$ singularity; (b) a representative scattering  diagram: the virtual $W$ propagator leads to a $1/V^2$ pole.
}\label{fignsc}
\end{figure}

For the process ${q\, X\rightarrow q'\, Y}$, considering the virtuality $V^2\equiv m^2-(P_q-P_q')^2$ defined in eq.~(\ref{virt}), we can divide the Feynman diagrams into two classes. The first class, that we name ``scattering diagrams'', consists of the diagrams with a $1/V^2$ pole (see figs.~\ref{scatt}, \ref{fignsc}b), while the second class consists of all the others. In particular, among various topologies, the second class contains  diagrams with $W$'s radiated from the $q q'$ fermion line and thus we   refer to it as the class of ``radiation diagrams'' , though the term is somewhat improper. An example of radiation diagram is shown in Fig.~\ref{fignsc}a, but other examples can be obtained from a scattering diagram like Fig.~\ref{fignsc}b, by permuting $q'$  with a quark in the inclusive final state $Y$. Notice that although the contribution of any given diagram is gauge dependent, the total contribution featuring a $1/V$ pole is obviously gauge independent.
Now, assessing the validity of the EWA amounts to assessing by how much and in which kinematic  regime the $1/V$ pole part in the scattering diagrams dominates over all the other contributions. In practice, by working in the axial gauge,  it will be enough to consider just the scattering diagrams and focus on their expansion in $\delta_\bot$ and $\delta_m$ in the kinematic region where $V\ll E$. In other words, the radiation diagrams neither imply further cancellations nor  bring in more sizeable corrections.
That is because of two reasons. First of all, as we shall see below,  in the axial gauge power counting is straightforward: unlike what happens in covariant gauges or in the unitary gauge, the amplitudes associated with individual Feynman diagrams do not feature spurious  $1/m$ singularities~\footnote{In those other gauges, the unphysical singular terms cancel when summing up all diagrams thus making power counting non-straightforward.}. Secondly, it is easy to see that, under the reasonable assumption that the quarks interact with the EWSB sector only via the exchange of $W$'s, the radiation diagrams cannot be enhanced with respect to the scattering ones by simply cranking up some coupling. That is obvious for diagrams that are obtained by permutation of quarks lines in scattering diagrams. 
Moreover inspection of  Fig.~\ref{fignsc} clarifies that is true also for diagrams with a genuine radiation structure: given one such diagrams in Fig.~\ref{fignsc}a we can construct a scattering one, Fig.~\ref{fignsc}b, by attaching the building blocks
 $W\rightarrow Y_i$ on a boson line.  The new diagram contains exactly the same couplings as the 
original one, meaning that any large coupling which is present in the 
radiation diagrams must unavoidably appear also in the 
scattering ones. Notice that the converse is not true, 
the simplest 
example is found in the higgsless SM where the strong quadrilinear 
coupling among longitudinal $W$'s, which grows with the energy as 
$(E/v)^2\gg1$, only contributes to $qW_L\rightarrow q'W_LW_L$ 
through a scattering diagram. In that situation, the radiation type
diagrams are further suppressed and it is even more justified 
to neglect them.
In view of the above we will not need to 
consider the radiation diagrams explicitly any longer, their effect 
is either comparable or smaller that that of  the subleading terms in the scattering diagrams.

\subsection{Splitting Amplitudes}
\label{splitting}

Let us now consider the generic weak process $qX\rightarrow q'Y$ defined 
in section~2.1, focusing  on the 
contributions from scattering diagrams of Fig.~\ref{scatt} to the amplitude
${\mathcal A}^{\textrm{sc}}(qX\rightarrow q'Y)$. We would like 
first of all to rewrite those diagrams, without any approximation, 
in a form suited to perform the expansion of eq.~(\ref{limit}). 
To this end we derive a decomposition the ${\mc N}$s in the vector boson propagator, for a generic 
 off-shell momentum $q$, in terms of a suitably chosen basis of polarization vectors, by 
proceeding as follows. A generic 4-momentum 
\beq
\displaystyle
q_\mu\,=\,\left(q_0,\,\vec{q}_\bot,\,q_L\right)_\mu\,,
\eeq
which we take for simplicity with  positive energy, but arbitrary norm 
$q^2=q_0^2-q_L^2-q_\bot^2$, can always 
be put in the form \footnote{The discussion which follows assumes that 
$q_0^2-q_\bot^2=q_L^2+q^2>0$, 
this condition obviously holds for the virtual $W$ momentum of eq.~(\ref{vwmom}) to 
which these manipulations will be applied.}
\beq
\displaystyle
{\overline{q}}_\mu\,=\,\left(\sqrt{q_L^2+q^2},\,\vec{0},\,q_L\right)_\mu\,,
\label{sf}
\eeq
by means of a Lorentz boost in a direction orthogonal  to the beam. Such a boost also leaves  
$n^\mu$ invariant. One can explicitly check that the boost characterized by the velocity 
$\vec{\beta}=(\vec{q}_\bot/q_0,0)$ does the job $$
B(\vec{q}_\bot/q_0,0)_\mu^{\;\;\nu}{\overline{q}}_\nu\,=\,q_\mu\,, \;\;\;\;\; B(\vec{q}_\bot/q_0,0)_\mu^{\;\;\nu}n_\nu\,=\,n_\mu\,,
$$
where $B(\vec{\beta})_\mu^{\;\;\nu}$ denotes the standard boost matrix.
Given that $n$ is invariant under $B$, it is convenient 
to work out the basis of polarization vectors 
in the boosted  frame where $q$ takes the form of eq.~(\ref{sf}),
and then obtain the ones in the original frame 
by boosting back with  $B$. In the boosted  frame the ${\mc N}_{\mu\nu}$ matrix takes a very simple diagonal 
form and it is immediate to identify the ``natural'' basis for its decomposition. Boosting it back we obtain
\bea
\displaystyle
&&{\mc N}_{\mu\nu}(q)\,=\,\sum_{h=\pm1}\left(\varepsilon_{\mu}^h\right)^*\varepsilon_{\nu}^h 
\,+\,\frac{1+q_L^2/m^2}{1+q_L^2/q^2}\,\varepsilon_{\mu}^0\varepsilon_{\nu}^0\,,\nn\\
&&{\mc N}_{\mu g}(q)\,=\,\varepsilon_{\mu}^0\,\varepsilon_g\,,\nn\\
&&{\mc N}_{g g}(q)\,=\,\varepsilon_{g}^*\,\varepsilon_g\,,
\label{dec}
\eea
where
\bea
\displaystyle
&&\varepsilon_\mu^{\pm1}(q)\,=\,
B(\vec{q}_\bot/q_0,0)_\mu^{\;\;\nu}\left(0,\,1/\sqrt{2},\,\mp i/\sqrt{2},\,0\right)_\nu\,,
\nn\\
&&\varepsilon_\mu^{0}(q)\,=\,
B(\vec{q}_\bot/q_0,0)_\mu^{\;\;\nu}\left(-\,\frac{\sqrt{1+q^2/q_L^2}}{\sqrt{1+q_L^2/m^2}},\,0,\,0,\,0\right)_\nu\,=\,-\,
\frac{m}{q_L\sqrt{1+m^2/q_L^2}}\left(n_\mu\,+\,\frac{q_\mu}{q_L}\right)\,,\nn\\
&&\varepsilon_g(q)\,=\,i\,\sqrt{1+m^2/q_L^2}\,.
\label{polv}
\eea
Clearly, the above equations rely on a choice of the normalizations and on the definition of 
the Goldstone's wave function $\varepsilon_g$, which we performed in such a way to get 
units coefficient in the decomposition (\ref{dec}) of  ${\mc N}_{\mu g}$, of  ${\mc N}_{ g g}$
 and of the transverse part of ${\mc N}_{\mu\nu}$.

In the on-shell limit $q^2\rightarrow m^2$ the numerator ${\mc N}_{IJ}$ reduces, as anticipated, 
to a rank-$3$ matrix and its $3$ non-null eigenvectors span the space of the physical $W$ states. 
Following \cite{Kunszt:1987tk}, we choose as a basis the two ``transverse'' wave-functions 
given by the  $5$-vectors $E_I^{\pm1}=(\varepsilon^{\mu\,\pm1},\,0)$, plus the ``longitudinal'' one 
$E_I^0=(\varepsilon^{\mu\,0},\,\varepsilon_g)$, and
${\mathcal N}_{IJ}=\sum_{h=\pm,0}\left(E_I^h\right)^*E_J^h$. 
Notice that with this definition 
the polarized $W$ states are 
not exactly eigenstates of the helicity but of the angular momentum 
along the beam direction 
in the $q=\overline{q}$ frame as it becomes apparent from eq.~(\ref{polv}) by checking that the polarization states transform with a phase under 
rotations in the plane transverse to the beam. In the lab frame, the transformation property of the polarization 
vectors under transverse rotations is also simple, let us work
 it out for future use.
Consider a rotation $R_\mu^{\;\;\nu}(\theta)$ of the momentum 
$q$ in the transverse plane, defined by
\beq
{q'}_\mu\rightarrow\,q_\mu^{(\theta)}\,=\,
R_\mu^{\;\;\nu}(\theta)q_\nu\,=\,(q_0,\,\cos\theta q_\bot^1+\sin\theta q_\bot^2,\,
\cos\theta q_\bot^2-\sin\theta q_\bot^1,\,q_L)_\mu\,.
\label{qrot}
\eeq
Tho composition rules of Lorentz transformations imply that
$$ 
B(\vec{q}_\bot^{(\theta)}/q_0,0)_\mu^{\;\;\nu}\,=\,R_\mu^{\;\;\mu'}B(\vec{q}_\bot/q_0,0)_{\mu'}^{\;\;\nu'}
{R^{-1}}_{\nu'}^{\;\;\nu}\,,
$$
from which it is immediate to check that the polarization 
vectors (\ref{polv}) satisfy
\beq
\displaystyle
{R^{-1}}_\mu^{\;\;\;\nu}\varepsilon_{\nu}^{h}(q^{(\theta)})\,=\,e^{ih\theta} 
\varepsilon_{\mu}^{h}(q)\,,
\label{pvtr}
\eeq
with $h=\pm1,\,0$. According to the above equation, which  is valid for a generic 
off-shell momentum $q$,  our polarization vectors correspond to wave functions with (total) angular momentum $h$ along the beam axis.

Let us now consider a generic scattering diagram, we will rewrite its amplitude by applying 
the decomposition (\ref{dec}). Notice that in the axial gauge, 
as Fig.~\ref{scatt} shows, there are  
two kind of scattering diagrams, which we denote as type ``A'' and ``B'' 
depending on whether the fermion line is attached to a 
gauge-gauge or to a gauge-Goldstone propagator. There is no contribution from the 
Goldstone-Goldstone propagator because  massless quarks  do 
not couple directly to the Goldstones. The momentum $K$ flowing in the 
$W$ propagator is given by eq.~(\ref{vwmom}), using eq.~(\ref{dec}) 
we rewrite each scattering diagram, 
up to corrections of ${\mc O}(\delta_\bot^2)+{\mc O}(\delta_m^2)$ 
that originate from expanding the second 
term in the first line of eq.~(\ref{dec}), as
\bea
\displaystyle
&&{\mc A}^{\textrm {sc.}-A}\,=\,-\frac{i}{V^2}\sum_{h=\pm1}\left[J^{\mu}
\left(\varepsilon^h_\mu\right)^*\right]
\left[\varepsilon^h_\nu {\mc A}_Q^{\nu}\right]\nn\\
&&\;\;\;\;\;\;\;\;\;\,-\,\frac{i}{V^2}\,
\left[J^{\mu}\left(\varepsilon^0_\mu\right)^*\right]
\left[\left(1\,-\,\frac{V^2}{m^2}\right)\varepsilon^0_\nu {\mc A}_Q^{\nu}
\right]\,\left[
1\,+\,{\mc O}(\delta_\bot^2+\delta_m^2)\right]
,
\nn\\
&&{\mc A}^{\textrm {sc.}-B}\,=\,-\,
\frac{i}{V^2}\,
\left[J^{\mu}\left(\varepsilon^0_\mu\right)^*\right]
\left[\varepsilon_s {\mc A}_Q^{g}\right]\,,
\label{scdec}
\eea
where ${\mc A}_Q^{\nu}$ denotes the ``hard'' amputated amplitude of the gauge field 
$W^{\nu}_Q$ on the $X$ and $Y$ 
states, while ${\mc A}_Q^{g}$ is the Goldstone amplitude on the same 
external states. 
The ``soft'' part of the scattering diagrams, associated to the quark vertex in eq.~(\ref{qint}), 
is instead contained in the chiral current $J$, which is given by
\beq
\displaystyle
J^{\mu}(P_{q'},\,P_q) \,\equiv\, -2iC\,\overline{u}_L(P_{q'})
\gamma^\mu u_L(P_q)\,=\,
-2iC\,\chi^\dagger(P_{q'})\overline{\sigma}^\mu \chi(P_q)\,,
\label{chcur}
\eeq
where $C$ depends on the charge of the virtual $W$ and is given in 
eq.~(\ref{Ci}).
The equation above has been written in the Weyl basis where the $\gamma$ matrices are
\beq
\gamma^\mu\,=\,
\left(
\begin{array}{cc}
0 & \sigma^\mu\\
\overline{\sigma}^\mu & 0
\end{array}\right)\,,\;\;\;\;\;
\gamma^5\,=\,
\left(
\begin{array}{cc}
-\I & 0\\
0 & \I
\end{array}\right)\,,\;\;\;\;\;
\eeq
and the wave function for massless spinors of helicity $-1/2$ and 
generic 
momentum $P_\mu\,=\,(p_0,\,p_1,\,p_2,\,p_3)_\mu$ can be taken to be
\beq
\displaystyle
u_L(P)\,=\,\left(
\begin{array}{c}
\displaystyle
\chi(P)\\
0
\end{array}\right)\,,\;\;\;\;\;
\chi(P)\,=\,\frac1{\sqrt{p_0+p_3}}\left(
\begin{array}{c}
p_0+p_3\\
p_1+i\,p_2 
\end{array}\right)\,.\;\;\;\;\;
\label{exsp}
\eeq
The  above choice for the spinor wave function is fully analogue to the choice of polarization vectors in eq.~(\ref{pvtr}):
it is obtained by boosting the helicity eigenfunction $\chi_0$ from the frame where $p_\perp=0$ to the lab frame
\beq
\chi(P)\,=\Lambda_L\chi_0\,=\,\Lambda_L(p_\perp/p_0)\left(
\begin{array}{c}
\sqrt{2p_3} \\
0
\end{array}\right)\,,\;\;\;\;\;
\label{boostfermion}
\eeq
where
\beq
\Lambda_L(p_\perp/p_0)=e^{-\eta_i\sigma_i /2}\qquad\qquad \eta_i=\frac{p_\perp^i}{2|p_\perp |}\ln\left (\frac{1-|p_\perp |/p_0}{1+|p_\perp |/p_0}\right )\, .
\eeq
In analogy with eq.~(\ref{pvtr}),  the spinor wave functions transform under rotations in the transverse plane according to
\beq
e^{-i\theta \sigma_3/2} \chi(p^{(\theta)})
=e^{-i\theta/2}\chi(p)\eeq
and thus correspond to eigenstates with $J_3=-1/2$. A consequence of the above is that the current, being helicity preserving,
 transforms as a vector under rotations in the transverse plane, without the 
appearance of extra momentum-dependent phases. 
\beq
\displaystyle
J^{\mu}({P'}^{(\theta)},\,P^{(\theta)}) \,=\,R^{\mu}_{\;\;\mu'}J^{\mu'}(P',\,P) \,.
\label{jtr}
\eeq

It is a simple exercise, at this point, to compute the ``splitting amplitudes'', {\it{i.e.}} the 
soft part of the amplitude (\ref{scdec}). In the expansion of eq.~(\ref{limit}) the result is
\bea
\displaystyle
&&-\frac{i}{V^2}\left[J^{\mu}\left(\varepsilon^{\pm1}_\mu\right)^*\right]\,=\,
2C
\frac{p_\bot e^{\pm i\phi}}{V^2}
\,g_\pm(x)\left[
1\,+\,{\mc O}(\delta_\bot^2+\delta_m^2)\right]
\,,
\nn\\
&&-\frac{i}{V^2}\left[J^{\mu}\left(\varepsilon^{0}_\mu\right)^*\right]\,=\,
2C\,
\frac{m}{V^2}
\,g_0(x)\left[
1\,+\,{\mc O}(\delta_\bot^2+\delta_m^2)\right]
\,,
\label{sf1}
\eea
where
\bea
\displaystyle
&&g_+(x)\,=\,\sqrt{2}\frac{\sqrt{1-x}}{x}\,,\nn\\
&&g_-(x)\,=\,\sqrt{2}\frac{1}{x\sqrt{1-x}}\,,\nn\\
&&g_0(x)\,=\,2\frac{\sqrt{1-x}}{x}\,.
\label{polf}
\eea
The dependence of the result on 
\begin{equation}
p_\bot e^{ -i\phi}\equiv p_\bot^1-ip_\bot^2\equiv \tilde p_\bot
\label{complexp}
\end{equation} 
follows from  eqs.~(\ref{pvtr}) and (\ref{jtr}), according to which
the splitting amplitude into a vector boson of helicity $h$ changes by a phase $e^{-i h\theta}$ under a rotation in the transverse plane. That result is a reflection of angular momentum conservation: the amplitude for helicity $h$ corresponds to a final state with orbital angular momentum $-h$ so as to compensate for the spin of the vector boson. By the resulting selection rule the longitudinal splitting amplitude arises at zeroth order in $p_\bot$ while the transverse splitting amplitudes 
only arise at first order. Similarly the subleading corrections to all amplitudes are quadratic in $p_\bot$. 
Notice that the subleading corrections  are also quadratic in the mass $m$, because $m$ appears linearly only  in the prefactor of the longitudinal 
splitting amplitude while all the other terms are quadratic. That result is  associated to a
reparametrization of the lagrangian and the fields under which: $W\to W$, $\pi \to -\pi$ and  $m\to -m$ (see discussion in section~\ref{correwa}).

\subsection{Derivation of  EWA}
\label{deriv}

In the previous section we worked out the fermion current factors in eq.~(\ref{scdec}),
which are associated with the collinear splitting $q\to q'W$.  Next we should deal with the factors associated with the hard transition. Those we would like to  replace with the 
on-shell $WX\rightarrow Y$ scattering amplitudes. The  hard amplitudes 
${\mc{A}}_Q^{\nu,\,g}$ 
and  the polarization vectors $\varepsilon^{h,\,g}$ depend on the momentum of the virtual $W$, parametrized by $K$ in eq.~(\ref{vwmom}). As $K$ is nearly on shell it
is convenient to rewrite it as
\beq
\displaystyle
K\,=\,\left(\omega_{\vec{k}}\sqrt{1-V^2/\omega_{\vec{k}}^2},\,\vec{k}\right)\,,
\label{wmom2}
\eeq
where $\omega_{\vec{k}}=\sqrt{m^2+|{\vec{k}}|^2}$ 
is the on-shell energy of a real $W$ with the same $3$-momentum 
${\vec k}=(-\vec{p}_\bot,xE)$ as the virtual $W$. The deviations 
from on-shellness only affect the energy component $K^0$ of $K$ by an amount of relative size
 ${\mc O}(V^2/\omega_{\vec{k}}^2)={\mc O}(\delta_\bot^2
+\delta_m^2)$. In the hard region the amplitudes are  well behaved functions of the external momenta
and so are the wave functions. Thus the subprocess amplitudes in of eq.~(\ref{scdec}) can  be safely expanded in the virtuality $V$
\beq
\left [\epsilon \cdot {\cal A}\right ]\equiv {\cal M}(V^2)={\cal M}(0)+{\cal M'}(0) V^2 + O(V^4)\,.
\eeq
The reasonable expectation is that, barring cancellations that we shall discuss below, the radius of convergence of the above expansion is controlled by the hardness of the subprocess, as  quantified by the virtuality $H$ of its internal lines, that is ${\cal M'}(0)/{\cal M}(0)\sim 1/H^2$. By assuming, for simplicity, $H\sim E$, we can
then write

\bea
\displaystyle
{\mc A}^{\textrm sc.-A}&&
\,=\,2C\,\frac{p_\bot}{V^2}\sum_{h=\pm1}e^{\mp i\phi}g_\pm(x)
\left[\varepsilon^+_\nu {\mc A}^{\nu}\right]_{\textrm{on}}\hspace{-4pt}(\vec{k})\,
\left[1+{\mc O}(\delta_\bot^2+\delta_m^2)
\right]\,\nn\\
&&+\,
2C\,\frac{m}{V^2}g_0(x)\left[\varepsilon^0_\nu {\mc A}^{\nu}\right]_{\textrm{on}}\hspace{-4pt}(\vec{k})
\left[1+{\mc O}(\delta_\bot^2+\delta_m^2)
\right]\,\nonumber\\
&&+2C \frac{1}{k_L}g_0(x)\,{\mc A}_{\textrm{local}}\,,
\nn\\
{\mc A}^{\textrm sc.-B}&&
\,=\,
2C\,\frac{m}{V^2}g_0(x)\left[
\varepsilon_g {\mc A}^{g}\right]_{\textrm {on}}\hspace{-4pt}(\vec{k})
\left[1+{\mc O}(\delta_\bot^2+\delta_m^2)
\right]\,,
\label{scat}
\eea
where all the amplitudes and the polarization vectors in the square 
brackets are computed, as explicitly indicated, for an  
on-shell $W$ with 4-momentum $(\omega_{\vec{k}},\vec{k})$. The term denoted as ${\mc A}^{\textrm{local}}$, collects all the 
contributions proportional to the $V^2/m^2$ term in the parenthesis in the second line 
of eq.~(\ref{scdec}), and is thus non singular as $V\to 0$.
Now, focusing on the interesting  regime $V\sim m \sim p_\bot$, it is instructive to power count the various terms. Those associated with the on-shell amplitudes are $O(1/V)$. The presence of a $1/V$ pole indicates that
the virtual W is emitted and absorbed at well separated spacetime points.  On the other hand the term ${\mc A}^{\textrm{local}}$ is $O(V^0)$ and not-singular as $V\to 0$, like the local amplitude mediated by a  contact term. Thus the term ${\mc A}^{\textrm{local}}$  belongs to the same class as the contribution 
from the diagrams with  radiation topology described in Section~\ref{sca}.
Explicitly we find
\beq
{\mc A}^{\textrm{local}}\,=\,\left(n_\nu
+K_\nu/k_L\right) {\mc A}_Q^{\nu}\left[1+{\mc O}(\delta_\bot^2+\delta_m^2)
\right]\,,
\label{lterm}
\eeq
where we made use of the explicit form of the $\varepsilon^0$ vector in 
eq.~(\ref{polv}). Finally, notice that  the  subleading corrections to the $1/V$ pole terms, those associated with $\delta_\bot^2$ and $\delta_m^2$ , correspond to  $O(V)$ contributions, that are thus further suppressed with respect to ${\mc A}^{\textrm{local}}$.

The above discussion would be invalidated by the occurrence of \emph{on-shell cancellations}, that 
badly change our estimate ${\cal M'}(0)/{\cal M}(0)\sim 1/H^2$. 
The most obvious thing that can happen is that ${\cal M}(0)=0$, corresponding to a vanishing amplitude for the subprocess, which is not our hypothesis~\footnote{For ${\cal M}(0)=0$ the $1/V$ pole is absent, corresponding  to the possibility of parametrizing the $qX\to q' Y$ process by a contact interaction. That can be done via a field redefinition or, equivalently, by using the $W$ equations of motion.}.
 Leaving out the trivial case ${\cal M}(0)=0$, we can  identify two other more subtle sources for such cancellations, one physical and  one unphysical. 
 The first possibility, corresponding to a genuine physical effects, is that ``helicity selection rules''~\footnote{Of the type involved in the cancellation of MHV amplitudes.} suppress ${\cal M}(0)$ by some power of $m/E$ with respect to naive power counting. This could happen for peculiar polarizations of the external states described by $X$ and $Y$, but we do not expect it to happen for all polarizations. Thus for sufficiently inclusive initial and final states there should be no issue: the estimate of the corrections in eq.~(\ref{scat}) will safely apply to the dominant helicity channels, which of course are the ones which are not subject to cancellations.  In section~\ref{wwsec} we will discuss examples of these these helicity induced cancellations. The second possibility, associated instead to an unphysical effect, arises in gauges where the propagator and/or the polarization vectors are singular for $m\to 0$. In those gauges, that include the unitary and $R_\xi$ gauges but not the axial gauge, naive power counting is spoiled by the presence of terms which, diagram by diagram,  grow off-shell like powers of $E/m$. More precisely, in those gauges one finds spurious contributions  ${\cal M'}(0)/{\cal M}(0)\sim E^2/m^4$ that  completely invalidate an approach based on diagrammatics \cite{Kleiss:1986xp}.
Those spurious terms are unphysical and cancel only upon summation of the whole set of diagrams.
The axial gauge  has the great advantage that the cancellation is manifest diagram by diagram,
simply because no such term can arise due to the absence of $1/m$ singularities!
The origin of this problem is simply that ${\cal M'}(0)$, unlike ${\cal M}(0)$, is an unphysical gauge dependent quantity.
Because of the above issue, the validity of 
 EWA was put in doubt by ref.~\cite{hep-ph/0608019}. Our basic point, that simply follows and extends the discussion in \cite{Kunszt:1987tk}, is that the problem is easily avoided by working in the axial gauge. In fact EWA can also conveniently be derived by working in the $R_\xi$ gauge, by noticing that one has the freedom of chosing a parametrization where the polarization vectors are non-singular as $m\to 0$. That is the parametrization that is most suitable for proving the equivalence between longitudinally polarized vectors and eaten Goldstones in high-energy scattering. We will present this alternative approach in a forthcoming paper.

In eq.~(\ref{scat}) we can already recognize the various terms of the 
EWA formula reported in eq.~(\ref{EWA}). At the leading 
order, the  terms in the first line in ${\mc A}^{sc.-A}$ coincide with the 
$W_QX\rightarrow Y$ scattering amplitude with transversely polarized 
incoming $W$'s while the term in second line, in combination with ${\mc A}^{sc.-B}$, 
reconstructs the longitudinally polarized amplitude. Indeed, in the  axial gauge,  the on-shell $W_L$ state is a coherent superposition of the gauge  and  Goldstone fields. That is why the  amplitude for $W_L$  involves  the sum of two  terms.  Using a compact notation the total $qX\rightarrow q'Y$ 
amplitude can therefore be rewritten as
\begin{equation}
\displaystyle
{\mc A}
\,=\,\frac{2C}{V^2}\left [ \tilde p_\bot
{\mc A}_{+}+\tilde p_\bot^*
{\mc A}_{-}+ m {\mc A}_{0}+ \frac{V^2}{k_L} {\mc A}_{\textrm{local}}\right ]\left[1+{\mc O}(\delta_m^2+\delta_\bot^2)\right ]\,,
\label{sclast}
\end{equation}
\begin{equation}
\displaystyle
{\mc A}_{\pm} = g_\pm(x)\left[\varepsilon^\pm_\nu {\mc A}^{\nu}\right]_{\textrm{on}}\hspace{-4pt}(\vec{k})\,,\qquad\qquad
{\mc A}_{0} =g_0(x)\left \{\left[\varepsilon^0_\nu {\mc A}_Q^{\nu}\right]_{\textrm{on}}\hspace{-4pt}(\vec{k})+
\left[\varepsilon_g {\mc A}^{g}\right]_{\textrm{on}}\hspace{-4pt}(\vec{k})\right \}\,.
\label{simple}
\end{equation}

Eq.~(\ref{sclast}), by neglecting the subleading
 ${\cal A}_{\textrm{local}}$ term (as well as the $\delta_\bot^2$ and $\delta_m^2$ corrections)
yelds a ``\emph{generalized} EWA''
\begin{equation}
\displaystyle
{\mc A}_{\textrm{gEWA}}
\,=\,\frac{2C}{V^2}\left [ \tilde p_\bot
{\mc A}_{+}+\tilde p_\bot^*
{\mc A}_{-}+ m {\mc A}_{0}\right ]
\label{gewa}
\end{equation}
which  provides an approximation
for the \emph{amplitude} of the complete process in terms of the hard on-shell 
scattering amplitude of an equivalent $W$ boson. The resulting differential cross section is more accurate 
than the standard approximation in  eq.~(\ref{EWA}), because it maintains the information on the  $\phi$  distribution 
of the forward jet, which  is instead integrated in the standard EWA. The $\phi$ distribution is determined by the interference among the
individual subamplitudes in eq.~(\ref{gewa}).
 Moreover, as we shall clarify  better below,  for processes dominated by ${\cal A}_0$, as it often happens at sufficiently high energy in composite Higgs models, this generalized EWA includes in a consistent manner the leading  $O(\delta_\bot)$ corrections to the differential rate coming from the interference between ${\cal A}_0$ and ${\cal A}_\pm$.
 We will discuss the implications of this generalized EWA formula in 
a forthcoming paper.

To obtain eq.~(\ref{EWA}) we just need to further expand the subamplitudes in $\frac{\tilde p_\bot}{E}, \frac{\tilde p_\bot^*}{E}$ 
\begin{equation}
{\mc A}_i= {\mc A}^{(0,0)}_i +{\mc A}^{(1,0)}_i\frac{\tilde p_\bot}{E} +{\mc A}^{(0,1)}_i \frac{\tilde p_\bot^*}{E}\dots
\label{expand}
\end{equation}
for $i=(+,-,0,{\textrm{local}})$ and keep only the very leading term
\begin{equation}
{\mc A}^{(0,0)}
\,=\,\frac{2C}{V^2}\left [ \tilde p_\bot
{\mc A}_{+}^{(0,0)}+\tilde p_\bot^* {\mc A}_{-}^{(0,0)}+ m {\mc A}_{0}^{(0,0)}\right ]
\label{amp11}
\end{equation}
The resulting approximation corresponds to replacing 
the momentum ${\vec k}=(-\vec{p}_\bot,xE)$ of the equivalent $W$, with a fully collinear one ${\vec k}_W=(-\vec{0},xE)$ as in eq.~(\ref{wmon}).
Modulo  the splitting function factor in front (see eq.~(\ref{simple})), $A^{(0,0)}_{+,-,0}$ represent the $W X\to Y$  amplitudes for   a fully collinear  equivalent $W$.
We thus find the differential cross section
\begin{eqnarray}
d\sigma(qX\rightarrow q'Y)_{\textrm{EWA}}\,=&&\,
\frac1{2E_q 2E_X|1-v_X|}\int_\phi\,\frac{|{\mc A}^{(0,0)}|^2}2\,
\frac{d^3 P_{q'}}{2E_{q'}\left(2\pi\right)^3}
\,d\Phi_Y(2\pi)^4\delta^4(P_Y^{\textrm tot}+P_{q'}-P_q-P_X)\nn\\
\,\simeq\,&&\frac{2C^2}{V^4}
\cdot \frac{p_\bot dp_\bot x dx}{(2\pi)^2 2(1-x)}\times \left [p_\bot^2
|{\mc A}_{+}^{(0,0)}|^2+ p_\bot^2|{\mc A}_{-}^{(0,0)}|^2+ m^2 |{\mc A}_{0}^{(0,0)}|^2\right ]
\nn\\
&&\times \frac1{2E_q 2E_W|v_W-v_X|} d\Phi_Y(2\pi)^4\delta^4(P_Y^{\textrm tot}-K_W-P_X)\,,
\label{EWA0}
\end{eqnarray}
where 
$d\Phi_Y$ denotes the phase space of the final state $Y$ 
and the $1/2$ factor comes from the average on the two polarizations 
of the incoming $q$. Notice that  by performing the $d\phi$ integral the interference terms cancel, since the ${\cal A}_i^{(0,0)}$ do not depend on $\phi$.
To obtain the second equality 
we have employed a few kinematic relations that are easily 
extracted from section~2.1, in particular we used that $E_q=E$, 
$d^3 P_{q'}=E p_\bot dp_\bot d\phi dx$ and that $E_{W}=xE$, $E_{q'}=(1-x)E$ 
up to quadratic corrections, we also used $v_W\simeq1$ in the relative velocity term. The incoherent sum of squared  amplitudes, when taking into account the splitting function coefficient in their definition eq.~(\ref{simple}),
together with the flux and phase space  factor for the $W X\to Y$ process in the third line, are then easily seen to reproduce eq.~(\ref{EWA}).
 
\subsection{Corrections to  EWA}
\label{correwa}

In the previous section we have established the validity of  EWA in the formal limit of extremely high energy, much above the jet $p_\bot$ and the $W$ mass. This result is reassuring as it guarantees the observability \emph{in principle} ({\it{i.e.}}, under ideal experimental conditions) of the on-shell $W$ boson collisions. In practice, however, the energy is limited and the EWA formula could receive large corrections. A hypothetical measurement of the equivalent $W$ boson scattering could therefore be affected by a potentially large intrinsic (systematic) error which is very important to quantify. This is the aim of the present section, in which we derive a parametric estimate of the deviations of the exact cross section from the EWA formula.

To start with, let us discuss the corrections due to the subleading terms in equation~(\ref{simple}), which we neglected in order to obtain the generalized EWA formula of  eq.~(\ref{gewa}). Aside the generic $\delta_\bot^2,\, \delta_m^2$ that originate from the expansion of the various matrix elements, corrections arise from the $ {\mc A}_{\textrm{local}}$ term, which is seemingly suppressed only by one power of  $\delta_\bot$, $\delta_m$ and therefore potentially gives the most sizable effect. However the relevance of $ {\mc A}_{\textrm{local}}$  crucially  depends on the  relative size of the amplitudes for different helicities ${\mc A}_{\pm},\,{\mc A}_{0},\ {\mc A}_{\textrm{local}}$. In general different sizes for the amplitudes involving the gauge fields on one side and the Goldstone bosons on the other are expected. That is  particularly true in models where the electroweak symmetry breaking sector is strongly coupled. Thus, while we expect ${\mc A}_{\pm}\sim {\mc A}_{\textrm{local}}$ as they both just involve the gauge field on the equivalent $W$ line, it turns out that ${\mc A}_{0}$ and $ {\mc A}_{\pm}$ typically have rather different sizes. In order to understand this point let us recall a useful selection rule controlling the appearance of powers of $m$ in physical quantities. The lagrangian for Goldstones and gauge fields (with or without a SM Higgs) is invariant under the reparametrization
\begin{equation}
W_\mu \to W_\mu\,, \qquad\qquad \pi\to -\pi\,, \qquad\qquad m\to - m\, ,
\label{symmetry}
\end{equation}
according to which the sign of $m$ is not a physical observable. Since the probability ${\mc A}{\mc A}^* $ is an observable, the amplitude ${\mc A}$ must be either even or odd under $m\to - m$. 
Indeed, because of eq.~(\ref{symmetry}) and because of  the structure of the polarization vectors $E^\pm_I, \,E^0_I$ (below eq.~(\ref{polv})), we conclude that
 ${\mc A}_\pm$ and  ${\mc A}_0$ have   opposite parities (more directly one can deduce that by noticing   the relative power of $m$ with which they enter ${\mc A}$).
In full generality  we must then have
\begin{equation}
\frac{{\mc A}_\pm}{{\mc A}_0}\,\sim\,\left (\frac{m}{E}\right )^{2n+1} b\,,
\label{ratiopm0}
\end{equation} 
where $b$ is a dimensionless ratio of couplings. In most cases (as in the examples of the following section) the simplest possibility is realized, that is either  $n=0$ or $n=-1$, so that the asymptotically  subleading polarized amplitude is suppressed  with respect to the leading one by $m/E$. In practice we therefore need to consider only two cases:
\begin{equation}
\textrm{1)}\;\; \frac{|{\mc A}_\pm|}{|{\mc A}_0|}\sim\frac{m}{E}b\,, \qquad \qquad\textrm{2)}\;\;\frac{|{\mc A}_0|}{|{\mc A}_\pm|}\sim\frac{m}{E}b\,,
\label{estt}
\end{equation}
(we stress that since the two cases are logically distinct,  the coefficient $b$ has  different meaning and size in the two cases).
From the above equations we see that in the typical situation there will be a hierarchy, due to the $m/E$ factor, among ${{\mc A}_\pm}$ and ${{\mc A}_0}$. It is however possible to obtain $|{{\mc A}_\pm}/{{\mc A}_0}|= O(1)$, but only in a specific range of energies and by the compensating effect of a very large $b$, of order $E/m$. We will illustrate an example of that in the following section, notice however that the by far more common situation is $b=O(1)$.

In order to better understand the nature of the $b$ parameter it is useful to restore dimensionality of $\hslash$. That way the coupling associated to an $n$-field vertex has dimensionality $n/2 - 1$. For instance the gauge coupling $g$ and the Higgs quartic $\lambda$ have respectively dimension $1/2$ and $1$. Since $b$ describes the ratio among homogeneous quantities it should be dimensionless. In any given theory we then know what powers to expect in $b$. For instance in the SM we must have $b=(g^2/\lambda)^p=(m^2/m_H^2)^p$. 

Let us consider now the squared amplitude integrated over $\phi$, which is the object we need in order to derive the final EWA formula in eq.~(\ref{EWA}). By expanding the subamplitudes  in $p_\bot$,   as in eq.~(\ref{expand}), under the reasonable assumption $E\partial_{p_\bot} {\mc A}_i\sim O({\mc A}_i)$, the result has the structure
\begin{equation}
\int_\phi |{\mc{A}}|^2\, \propto\,  p_\bot^2 {\mc A}_\pm^2+ m^2{\mc A}_0^2+{p_\bot^2}\frac{m}{E}{\mc A}_\pm{\mc A}_0
+V^2\frac{m}{E} {\mc A}_\pm{\mc A}_0+\frac{V^4}{E^2}{\mc A}_\pm^2+\dots\, ,
\label{integrated}
\end{equation}
where to simplify the notation we simply indicated by ${\mc A}_\pm$ and ${\mc A}_0$ the leading  terms
${\mc A}_\pm^{(0,0)}$ and    ${\mc A}_0^{(0,0)}$, defined in  eq.s~(\ref{expand},\ref{amp11}), and estimated the subleading terms according to ${\mc A}_i^{(m,n)}\sim {\mc A}_i$.
 In eq.~(\ref{integrated}), the third term arises from cross terms of the type $({\mc A}_0^{(0,0)})^*{\mc A}_+^{(0,1)}$ and $({\mc A}_0^{(1,0)})^*{\mc A}_+^{(0,0)}$, while the fourth term comes from $({\mc A}_0^{(0,0)})^*{\mc A}_{\textrm{local}}^{(0,0)}$. The dots represent terms of even higher order. Notice that the leading interference between ${\mc A}_\pm$ and ${\mc A}_{\textrm{local}}$ vanishes upon integration over $\phi$.
Focussing again on the physically interesting region $p_\bot\sim m$ we find that the relative corrections to the EWA formula scale like
\begin{equation}
\delta_{\textrm{EWA}}\,\sim\,
 \frac{m}{E}\frac{|{\mc A}_\pm{\mc A}_0|}{\max (|{\mc A}_\pm|^2,|{\mc A}_0|^2)}\sim \frac{m}{E}\min(|{\mc A}_0/{\mc A}_\pm|,|{\mc A}_\pm/{\mc A}_0|)\,.
\label{dEWA}
\end{equation}
We see that $\delta_{\textrm{EWA}}$ is always smaller than $m/E$, and it become of order $m/E$ only if ${\mc A}_0$ and ${\mc A}_\pm$ are comparable. 

In both cases considered in eq.~(\ref{estt}) $\delta_{\textrm{EWA}}$ becomes
\begin{equation}
\displaystyle
\delta_{\textrm{EWA}}\,\sim\,\frac{m}{E}\min\left[\frac{mb}{E},\frac{E}{mb}\right]\,.
\label{ddd}
\end{equation}
Let us analyze this formula in some more detail. For asymptotically high energies,  $\delta_{\textrm{EWA}}$ does scale quadratically with the energy, but with a possibly large prefactor $b$: $\delta_{\textrm{EWA}}\sim b m^2/E^2$. The corrections grow as $E$ decreases and at the critical value $E\simeq mb$ they become of order $m/E\sim1/b$. This behavior could be captured by a phenomenological formula

\begin{equation}
\displaystyle
\delta_{\textrm{EWA}}\,=\,C_1\frac{m^2}{E^2}b\frac{1}{1+C_2\frac{m^2}{E^2} b^2}\,,
\label{ddd1}
\end{equation}
with $C_{1,2}$ order one parameters. 
Notice that in the above discussion we have been implicitly assuming $b>1$. Only for  that case, does eq.~(\ref{ddd1})  represent the leading correction. Otherwise, $\delta_{\textrm{EWA}}$ is dominated by  
 $\delta_\bot^2,\, \delta_m^2$ in equation~(\ref{simple}), that is the the irreducible effects of taking the on-shell limit of the various matrix elements. In the end we always have $\delta_{\textrm{EWA}}>m^2/E^2$.
 
One region worth considering is $p_\bot\ll m$, focussing for simplicity on the more common case $b\simeq1$. In that case, the corrections scale differently for processes dominated by the longitudinal or the transverse amplitude, which correspond respectively to case 1) and case 2)   in eq.~(\ref{estt}). In  case 1),  eq.~(\ref{integrated}) implies as usual
\begin{equation}
\displaystyle
\delta_{\textrm{EWA}}\,\sim\,\frac{m^2}{E^2}\,.
\end{equation}
In case 2), instead, the corrections from the local term are enhanced
\begin{equation}
\displaystyle
\delta_{\textrm{EWA}}\,\sim\,\frac{m^2}{E^2}\frac{m^2}{p_\bot^2}\,,
\end{equation}
notice in particular that the EWA breaks down completely for $p_\bot \sim m^2/E$. For smaller $p_T$, the transverse contribution becomes negligible, while the contribution from ${\mc A}_0\sim (m/E) {\mc A}_\pm$ scales precisely like the  local contribution.
Notice that a similar phenomenon does not take place for the splitting into massless vector bosons, like gluons and photons, since in that case the virtuality $V^2$ also  goes to zero  like $p_\bot^2$.

Finally, we should also consider the case of intermediate jet transverse momentum:  $E\gg p_\bot\gg m$, in which the corrections are always
\begin{equation}
\displaystyle
\delta_{\textrm{EWA}}\,\sim\,\frac{p_\bot^2}{E^2}\,.
\end{equation}
Notice that in the more typical situation, $b\sim1$ and $p_\bot\not\ll m$, the corrections are simply given, as anticipated in section~\ref{BasicPicture}, by eq.~(\ref{EWAcorr}).

\subsection{The example of $\mathbf{WW}$ scattering}
\label{wwsec}

The general considerations of the previous sections are conveniently illustrated in the explicit example of the $WW$ scattering process, which we already considered in section~\ref{nuns}. We work in the Higgs model (with vanishing hypercharge, compatibly with eq.~(\ref{qint})) and we compute the totally polarized on-shell scattering amplitude $W^+_{p_1}W^-_{p_2}\to W^+_{p_3}W^-_{p_4}$. For instance, in the high energy ($E\gg m$, $E\gg m_H$) and fixed angle limit, the 
{${++++}$} amplitude ({\it{i.e.}}, $\{p_1,p_2,p_3,p_4\}=\{1,1,1,1\}$), reads
\begin{equation}
\displaystyle
{\mc A}_+=-\frac{2g^2}{\sin^2{(\theta/2)}}+{\mc O}(m^2/E^2)\,,
\label{piu}
\end{equation}
where $E$ and $\theta$ denote respectively  the energy and the scattering angle of the $W^+$ in the center of mass frame. The above result complies perfectly with the expectations of power counting for the scattering amplitude among transversely polarized $W$'s: two powers of $g$ and constant scaling  with energy. Moreover  the Coulomb singularity at $\theta=0$ is due to the negligibility of the $W$ mass  in the limit of high-energy and fixed angle. Consider now, instead, the {${-+++}$} amplitude, we would expect the same scaling with the energy but instead we find 
\begin{equation}
\displaystyle
{\mc A}_-=\frac{m^2}{E^2}\cdot\frac{3g^2\cos^2{(\theta/2)}}{2}\,,
\label{menoo}
\end{equation}
which is suppressed by an additional factor $m^2/E^2$. This is one example of the ``helicity-induced'' cancellations we mentioned in section~\ref{deriv}: the \emph{on-shell} amplitude is anomalously reduced with respect to its power counting estimate and there is no reason why this reduction should persist also in the case of off-shell external states. This potentially constitutes a problem for our derivation of the EWA formula because it could lead to an enhancement of the relative corrections, as explained in section~\ref{deriv}. Actually, the cancellation 
in equation~(\ref{piu}) could have been guessed by remembering the usual helicity cancellations in massless gauge theories. By crossing the {${----}$} amplitude and making all the external lines initial we obtain {${-+++}$}, since the latter must vanish in the massless limit, there should be a reduction for $m\ll E$. Notice that this reduction factor must be an even power of $m/E$, because of the $m\to-m$ selection rule of eq.~(\ref{symmetry}). Again, because of that selection rule, the {${0+++}$} amplitude must be odd under $m\to-m$ and indeed we find
\begin{equation}
\displaystyle
{\mc A}_0=\frac{m}{E}\cdot\frac{g^2\cot{(\theta/2)}}{\sqrt{2}}\,.
\label{zero}
\end{equation}
The suppression in this case is perfectly understood by power-counting, therefore it must persist also off-shell and does not signal any worrisome on-shell cancellation.

However the helicity induced cancellation which we discovered in eq.~(\ref{menoo}) is not very dangerous. In particular it does not invalidate our derivation of the EWA for the process $qW^-_{p_3}\to q' W^+_{p_3} W^-_{p_3}$, with $\{p_1,p_2,p_3\}=\{1,1,1\}$, where the external helicities are  {${+++}$}.  In that case, all the three amplitudes ${\mc A}_\pm$ and ${\mc A}_0$  appear, corresponding to the possible helicities of the intermediate equivalent $W$. The leading contribution to the total $2\to3$ amplidute comes from just ${\mc A}_+$ in eq.~(\ref{piu}) while the others, and in particular ${\mc A}_-$, are subleading precisely because they are canceled. The on-shell cancellation simply implies that we cannot control the off-shell corrections to ${\mc A}_-$, but these are irrelevant because they are at most as big as ${\mc A}_-$, and anyhow subleading with respect to ${\mc A}_+$. In practice, the derivation is saved by the ${\mc A}_+$ term, which is leading and not canceled. Consider now instead the {${-++}$} process, $\{p_1,p_2,p_3\}=\{-1,1,1\}$, for which the three sub-amplitudes read
\begin{eqnarray}
&&{\mc A}_+=\frac{m^2}{E^2}\cdot\frac{3g^2\cos^2{(\theta/2)}}{2}\,,\nonumber\\
&&{\mc A}_0=\frac{m^3}{E^3}\cdot\frac{g^2\left(-4+m_H^2/m^2+9\cos\theta\right)}{8\sqrt{2}\tan{(\theta/2)}}\,,\nonumber\\
&&{\mc A}_-=\frac{m^4}{E^4}\cdot\frac{g^2\left(2m_H^2/m^2+9\cos\theta-1/2(3+\cos\theta)\csc^2{(\theta/2)}\right)}{16}\,.
\label{222}
\end{eqnarray}
The complicated trigonometric structure of the above equations and also the dependence on the Higgs-$W$ mass ratio $m_H/m$, which we take momentarily to be of order one, do not play any role in the following discussion, for which we just need to focus on the $m/E$ dependence of the various terms. Denoting $\varepsilon=m/E$, we see that ${\mc A}_+$ and ${\mc A}_-$ scale respectively as $\varepsilon^2$ and $\varepsilon^4$ while they were expected from power counting to be of order $\varepsilon^0$. The suppression of both ${\mc A}_+$ and ${\mc A}_-$ can again be understood in terms of the amplitude cancellations in massless gauge theories, after crossing, these amplitudes become respectively {${-+--}$}and {${----}$}. The occurrence of the \emph{double} cancellation in ${\mc A}_-$ is instead unexplained~\footnote{Notice that also the longitudinal amplitude ${\mc A}_0$ is suppressed in eq.~(\ref{222}), it scale like $\varepsilon^3$ instead than $\varepsilon$. This cancellation cannot be inferred by the standard MHV cancellations in the massless theory because it involves the longitudinal polarizations. It corresponds, in the massless theory, to a cancellation of the Goldstone amplitude with three gauge fields of {${---}$} helicity (after crossing).}. For what concerns the validity of the EWA, the problem with the {${-++}$} process is that \emph{both} the polarized subamplitudes that are expected to lead  by power counting are canceled! Because of that cancellation we cannot control the size of the corrections, which could, in principle, be as big as the leading order, thus invalidating the EWA. Of course, the price of the multiple cancellation is that the \emph{total} amplitude is also canceled so that the potential failure of the EWA in this particular helicity channel, which is subleading, would not show up as a violation of the EWA if we compute only the unpolarized cross section. The dominant helicity amplitudes, such as the {${+++}$} previously discussed, are by definition not canceled, and the derivation of the EWA applies without caveats. 

The message is then clear: our derivation of EWA, as we presented it in the previous sections, does
apply to polarized
processes where at least one subamplitude is not suppressed, and, a fortiori, for the total unpolarized cross section. However, at least in its present formulation, our proof does
 not apply to polarized processes where a cancellation occurs in all intermediate channels. Looking at table~\ref{TABL}, where we collected the scaling with $\varepsilon=m/E$ of all the helicity amplitudes, we find that, up to parity and charge conjugation, the polarized processes where our proof fails  are {${-++}$}, {${0++}$}, {${00+}$} and {${-0+}$}. The same caveat could apply, a priori, also to the $q{\overline{q}}\to q' {\overline{q}}'W^+_{p_1} W^-_{p_2}$ polarized $2\to4$ processes. However, as table~\ref{TABL} shows, it never happens that \emph{all} the helicity channels associated to the two equivalent $W$'s cancels simultaneously, there is always at least one channel which is not canceled, i.e. of order $\varepsilon^0$.

We stress that the above discussion only shows that there our derivation is not valid in some case, and not that the EWA must necessarily fail. It is not excluded that some other mechanism,  not taken into account in our approach, like for instance a cancellation affecting the total (exact) $2\to3$ amplitude, suppresses also the corrections by the same amount as the on-shell EWA amplitudes. As a matter of fact, the numerical checks performed in section~\ref{nuns} show that must indeed be the case! The graphs displayed in that section show that  EWA is satisfied also by those $2\to3$ helicity channels where the leading sub-amplitude is suppressed. We find this rather interesting because it means that the diagrammatic methods employed in the previous section, in which the origin of helicity induced cancellations is not transparent, does not capture entirely the essence of the EWA and the reasons for its validity. This suggests that it might be worth looking for a \emph{non-diagrammatic} proof of the EWA, which would encompass also the ``anomalous'' processes.

\begin{table}
\begin{tabular}{rrc}
	\multicolumn{3}{c}{$W_+^{\textrm{out}},\,W_-^{\textrm{out}}=+\,+$}	\\
	\cline{1-3}
	$W_+^{\textrm{in}}$	&$W_-^{\textrm{in}}$ &Scaling 					\\
	$+$				&$+$			&$\varepsilon^0$					\\
	$0$				&$+$			&$\varepsilon$					\\
	$-$				&$+$			&$\varepsilon^2$					\\
	$+$				&$0$			&$\varepsilon$					\\
	$0$				&$0$			&$\varepsilon^2$					\\
	$-$				&$0$			&$\varepsilon^3$					\\
	$+$				&$-$			&$\varepsilon^2$					\\
	$0$				&$-$			&$\varepsilon^3$					\\
	$-$				&$-$			&$\varepsilon^4$
\end{tabular}
\hspace{-3pt}
\begin{tabular}{rrc}
	\multicolumn{3}{c}{$ W_+^{\textrm{out}},\,W_-^{\textrm{out}}= 0\,0 $}	\\
	\cline{1-3}
	$W_+^{\textrm{in}}$	&$W_-^{\textrm{in}}$ &Scaling 					\\
	$+$				&$+$			&$\varepsilon^2$					\\
	$0$				&$+$			&$\varepsilon$					\\
	$-$				&$+$			&$\varepsilon^0$					\\
	$+$				&$0$			&$\varepsilon$					\\
	$0$				&$0$			&$\varepsilon^0$					\\
	$-$				&$0$			&$\varepsilon$					\\
	$+$				&$-$			&$\varepsilon^0$					\\
	$0$				&$-$			&$\varepsilon$					\\
	$-$				&$-$			&$\varepsilon^2$
\end{tabular}
\hspace{-3pt}
\begin{tabular}{rrc}
	\multicolumn{3}{c}{$W_+^{\textrm{out}},\,W_-^{\textrm{out}}=+,\,-$}	\\
	\cline{1-3}
	$W_+^{\textrm{in}}$	&$W_-^{\textrm{in}}$ &Scaling 					\\
	$+$				&$+$			&$\varepsilon^2$					\\
	$0$				&$+$			&$\varepsilon$					\\
	$-$				&$+$			&$\varepsilon^0$					\\
	$+$				&$0$			&$\varepsilon$					\\
	$0$				&$0$			&$\varepsilon^0$					\\
	$-$				&$0$			&$\varepsilon$					\\
	$+$				&$-$			&$\varepsilon^0$					\\
	$0$				&$-$			&$\varepsilon$					\\
	$-$				&$-$			&$\varepsilon^2$
\end{tabular}
\hspace{-3pt}
\begin{tabular}{rrc}
	\multicolumn{3}{c}{$W_+^{\textrm{out}},\,W_-^{\textrm{out}}=0\,+$}	\\
	\cline{1-3}
	$W_+^{\textrm{in}}$	&$W_-^{\textrm{in}}$ &Scaling 					\\
	$+$				&$+$			&$\varepsilon$					\\
	$0$				&$+$			&$\varepsilon^0$					\\
	$-$				&$+$			&$\varepsilon$					\\
	$+$				&$0$			&$\varepsilon^2$					\\
	$0$				&$0$			&$\varepsilon$					\\
	$-$				&$0$			&$\varepsilon^2$					\\
	$+$				&$-$			&$\varepsilon$					\\
	$0$				&$-$			&$\varepsilon^2$					\\
	$-$				&$-$			&$\varepsilon^3$
\end{tabular}
\caption{The table shows the scaling of the polarized amplitudes with the parameter $\varepsilon=m/E$ for $\varepsilon\rightarrow0$. In the limit, the Higgs mass $m_H$ is kept constant and of order $m$. The missing combinations can be obtained by exploiting the $C$ and $P$ symmetry of the $W$ lagrangian.}
\label{TABL}
\end{table}

Finally, we would like to use $WW$ scattering to illustrate the possible enhancement 
of the subleading corrections via a large ratio of couplings $b$, as discussed in the previous section. 
 Consider the {${0+0}$} process, the subamplitudes are
\begin{eqnarray}
&&{\mc A}_+=g^2\cot^2{(\theta/2)}\,,\nonumber\\
&&{\mc A}_0=-\frac{m}{E}\cdot\frac{g^2\left(2+m_H^2/m^2+3\cos\theta\right)}{4\sqrt{2}\tan{(\theta/2)}}\,,\nonumber\\
&&{\mc A}_-=\frac{m^2}{E^2}\cdot\frac{g^2\left(4+m_H^2/m^2+9\cos\theta\right)}{8}\,.
\label{333}
\end{eqnarray}
Apart from the by now habitual $m^2/E^2$ cancellation which affects ${\mc A}_-$, the peculiarity of this channel is  that the Higgs quadrilinear coupling $\lambda$, which appears through the ratio $m_H^2/m^2\simeq\lambda/g^2\equiv b$, does not contribute to all the subprocesses. In particular, it does not contribute to the leading process ${\mc A}_+$. Therefore, if $b\gg 1$, the subleading process ${\mc A}_0$ is enhanced and this results in an enhancement of the relative corrections as explained in section~\ref{correwa}. Equation~(\ref{estt}) reads, in this particular case, 
\begin{equation}
\frac{{\mc A}_0}{{\mc A}_+}\simeq \frac{m}{E}\,b\, = \frac{m_H^2}{Em}\,,
\end{equation}
which  could be made parametrically large in the hypothetical situation of a very light $W$ boson. 

\section{Conclusions and Outlook}

In this paper we have discussed how, in an ideal experimental situation,  it is possible to access the scattering process of on-shell equivalent $W$ bosons disentangling it from the complete partonic interaction. We have shown that this can be achieved in the kinematic regime of forward energetic jets, where the complete process factorizes as a hard scattering convoluted with the collinear emision of the equivalent $W$ boson. This is of course nothing but a statement on the validity of  EWA, which we have found to hold up to corrections that scale quadratically with the hardness $H$ of the $W$ interaction. Quantitatively, the relative deviations can be typically estimated as {Max$[p_\bot^2/H^2,\,m^2/H^2]$}, with $p_\bot$ the forward jet transverse momentum. Actually, we have also found that the corrections could be enhanced, but only in very peculiar situations with, probably, a very limited practical impact.

Our work could be extended in several directions, some of which we will explore in a forthcoming publication. First of all, we should quantify better the level of accuracy of EWA in the practical experimental conditions of the LHC collisions. Notice that the deviations from EWA are regarded, from our viewpoint, as \emph{systematic errors} in the determination of the equivalent $W$ boson cross section, thus it is crucial that they  be kept under control. On the theoretical side, we plan to complement the analysis of the present paper, based on the axial gauge, with a covariant gauge derivation of  EWA which presents several interesting aspects. It would also be interesting, in the future, to extend the derivation to higher orders in the perturbative expansion, including in the first place  QCD radiative corrections. Another result of our paper deserving further study is the derivation of a generalized EWA formula,  eq.~(\ref{gewa}), which provides a prediction of the totally differential cross section. In  particular also describes the distribution of the forward jet azimuthal angle, which is instead integrated over in the standard EWA.

Finally, one interesting aspect of our derivation is that it is definitely not complete because it does not account for the validity of  EWA in some peculiar polarized processes, like those listed in section~\ref{wwsec}, which are affected by on-shell suppressions closely analog to the cancellation of tree-level polarized amplitudes in massless gauge theories. Since the latter ones do not have a clear interpretation in terms of Feynman diagrams, it is not surprising that dealing with such processes becomes cumbersome with our diagrammatic methods. To go beyond, probably, a non-diagrammatic approach would be needed.

\section*{Acknowledgments}

We thank C.~Anastasiou, A.~Ballestrero, A.~Banfi, R.~Contino, S.~Frixione, P.~Lodone,  Z.~Kunszt and D.~Zeppenfeld for useful discussions. RF thanks Thomas Hahn for his help on the use of the \textsc{Cuba} library. This research is partially supported by the Swiss National Science Foundation under grants 200020-126941 and 200020-138131.  The work of RF is also partly supported  by the NSF under grants PHY-0910467 and PHY-0652363, and by the Maryland Center for Fundamental Physics. 
This research was supported in part by the European Programme Unification in the LHC Era, contract PITN-GA-2009-237920 (UNILHC) and by the ERC Advanced Grant no.267985 Electroweak Symmetry Breaking, Flavour and Dark Matter: One Solution for Three Mysteries (DaMeSyFla).

%%%%%%%%%%%%%%%


\begin{thebibliography}{99}
%
\bibitem{Chanowitz} 
  M.~S.~Chanowitz and M.~K.~Gaillard,
  %``The TeV Physics of Strongly Interacting W's and Z's,''
  Nucl.\ Phys.\ B {\bf 261}, 379 (1985).
  %%CITATION = NUPHA,B261,379;%%
  M.~S.~Chanowitz and M.~K.~Gaillard,
  %``Multiple Production of W and Z as a Signal of New Strong Interactions,''
  Phys.\ Lett.\ B {\bf 142}, 85 (1984).
  %%CITATION = PHLTA,B142,85;%%


   J.~Bagger, V.~D.~Barger, K.~-m.~Cheung, J.~F.~Gunion, T.~Han, G.~A.~Ladinsky, R.~Rosenfeld and C.~-P.~Yuan,
  %``CERN LHC analysis of the strongly interacting W W system: Gold plated modes,''
  Phys.\ Rev.\ D {\bf 52}, 3878 (1995)
  [hep-ph/9504426].
  %%CITATION = HEP-PH/9504426;%%
  J.~Bagger, V.~D.~Barger, K.~-m.~Cheung, J.~F.~Gunion, T.~Han, G.~A.~Ladinsky, R.~Rosenfeld and C.~P.~Yuan,
  %``The Strongly interacting W W system: Gold plated modes,''
  Phys.\ Rev.\ D {\bf 49}, 1246 (1994)
  [hep-ph/9306256].
  %%CITATION = HEP-PH/9306256;%% 

  
    C.~Englert, B.~Jager, M.~Worek and D.~Zeppenfeld,
  %``Observing Strongly Interacting Vector Boson Systems at the CERN Large Hadron Collider,''
  Phys.\ Rev.\ D {\bf 80}, 035027 (2009)
  [arXiv:0810.4861 [hep-ph]].
  %%CITATION = ARXIV:0810.4861;%%
    V.~D.~Barger, K.~-m.~Cheung, T.~Han and D.~Zeppenfeld,
  %``Single forward jet tagging and central jet vetoing to identify the leptonic $W W$ decay mode of a heavy Higgs boson,''
  Phys.\ Rev.\ D {\bf 44}, 2701 (1991)
  [Erratum-ibid.\ D {\bf 48}, 5444 (1993)].
  %%CITATION = PHRVA,D44,2701;%%

  
    A.~Ballestrero, D.~B.~Franzosi and E.~Maina,
  %``Vector-Vector scattering at the LHC with two charged leptons and two neutrinos in the final state,''
  JHEP {\bf 1106}, 013 (2011)
  [arXiv:1011.1514 [hep-ph]].
  %%CITATION = ARXIV:1011.1514;%%
  A.~Ballestrero, G.~Bevilacqua and E.~Maina,
  %``A Complete parton level analysis of boson-boson scattering and ElectroWeak Symmetry Breaking in lv + four jets production at the LHC,''
  JHEP {\bf 0905}, 015 (2009)
  [arXiv:0812.5084 [hep-ph]].
  %%CITATION = ARXIV:0812.5084;%%
  
      J.~M.~Butterworth, B.~E.~Cox and J.~R.~Forshaw,
  %``$W W$ scattering at the CERN LHC,''
  Phys.\ Rev.\ D {\bf 65}, 096014 (2002)
  [hep-ph/0201098].
  %%CITATION = HEP-PH/0201098;%%
  
  
  K.~Doroba, J.~Kalinowski, J.~Kuczmarski, S.~Pokorski, J.~Rosiek, M.~Szleper and S.~Tkaczyk,
  %``The W_L W_L scattering at the LHC: improving the selection criteria,''
  arXiv:1201.2768 [hep-ph].
  %%CITATION = ARXIV:1201.2768;%%  
  
  
``Vector Boson Scattering at High Mass'' in G.~Aad {\it et al.}  [The ATLAS Collaboration],
  %``Expected Performance of the ATLAS Experiment - Detector, Trigger and Physics,''
  arXiv:0901.0512 [hep-ex].
  %%CITATION = ARXIV:0901.0512;%%
  
  
%\cite{Contino:2010mh}
\bibitem{Contino:2010mh} 
  R.~Contino, C.~Grojean, M.~Moretti, F.~Piccinini and R.~Rattazzi,
  %``Strong Double Higgs Production at the LHC,''
  JHEP {\bf 1005}, 089 (2010)
  [arXiv:1002.1011 [hep-ph]].
  %%CITATION = ARXIV:1002.1011;%%
  
 \bibitem{equivalentphoton}
 E.~Fermi,
 %``On the theory of collisions between atoms and electrically charged particles,''
 Nuovo Cim.\ , 2,143 (1925)
 [hep-th/0205086].
 %%CITATION = HEP-TH/0205086;%%
 E.~J.~Williams,
 %``Nature of the high-energy particles of penetrating radiation and status of ionization and radiation formulae,''
 Phys.\ Rev.\  {\bf 45}, 729 (1934).
 %%CITATION = PHRVA,45,729;%%
 C.~F.~von Weizsacker,
 %``Radiation emitted in collisions of very fast electrons,''
 Z.\ Phys.\  {\bf 88}, 612 (1934).
 %%CITATION = ZEPYA,88,612;%%
 
   %\cite{Dawson:1984gx}
\bibitem{Dawson:1984gx} 
  S.~Dawson,
  %``The Effective W Approximation,''
  Nucl.\ Phys.\ B {\bf 249} (1985) 42.
  %%CITATION = NUPHA,B249,42;%%
 G.~L.~Kane, W.~W.~Repko and W.~B.~Rolnick,
  %``The Effective W+-, Z0 Approximation for High-Energy Collisions,''
  Phys.\ Lett.\ B {\bf 148}, 367 (1984);
  %%CITATION = PHLTA,B148,367;%%
  
\bibitem{EWAbulk} 
 R.~N.~Cahn and S.~Dawson,
 %``Production of Very Massive Higgs Bosons,''
 Phys.\ Lett.\ B {\bf 136}, 196 (1984)
 [Erratum-ibid.\ B {\bf 138}, 464 (1984)].
 %%CITATION = PHLTA,B136,196;%%
 R.~N.~Cahn,
 %``Production of Heavy Higgs Bosons: Comparisons of Exact and Approximate Results,''
 Nucl.\ Phys.\ B {\bf 255}, 341 (1985)
 [Erratum-ibid.\ B {\bf 262}, 744 (1985)].
 %%CITATION = NUPHA,B255,341;%%
 G.~Altarelli, B.~Mele and F.~Pitolli,
 %``Heavy Higgs Production at Future Colliders,''
 Nucl.\ Phys.\ B {\bf 287}, 205 (1987).
 %%CITATION = NUPHA,B287,205;%%
 J.~F.~Gunion, J.~Kalinowski and A.~Tofighi-Niaki,
 %``Exact F F ---> F F W W Calculation For The Charged Current Sector And Comparison With The Effective W Approximation,''
 Phys.\ Rev.\ Lett.\  {\bf 57}, 2351 (1986).


  J.~Lindfors,
  %``Distribution Functions For Heavy Vector Bosons Inside Colliding Particle Beams,''
  Z.\ Phys.\ C {\bf 28}, 427 (1985).
  %%CITATION = ZEPYA,C28,427;%%
  P.~W.~Johnson, F.~I.~Olness and W.~-K.~Tung,
  %``The Effective Vector Boson Method For High-energy Collisions,''
  Phys.\ Rev.\ D {\bf 36}, 291 (1987).
  %%CITATION = PHRVA,D36,291;%%
  I.~Kuss and H.~Spiesberger,
  %``Luminosities for vector boson - vector boson scattering at high-energy colliders,''
  Phys.\ Rev.\ D {\bf 53}, 6078 (1996)
  [hep-ph/9507204].
  %%CITATION = HEP-PH/9507204;%%





%\cite{Kleiss:1986xp}
\bibitem{Kleiss:1986xp}
  R.~Kleiss, W.~J.~Stirling,
  %``Anomalous High-energy Behavior In Boson Fusion,''
  Phys.\ Lett.\  {\bf B182 } (1986)  75.
 
 


%\cite{hep-ph/0608019}
\bibitem{hep-ph/0608019} 
  E.~Accomando, A.~Ballestrero, A.~Belhouari and E.~Maina,
  %``Isolating Vector Boson Scattering at the LHC: Gauge cancellations and the Equivalent Vector Boson Approximation vs complete calculations,''
  Phys.\ Rev.\ D\ {\bf 74}, 073010  (2006)
  [hep-ph/0608019].
  %%CITATION = PHRVA,D74,073010;%%
      A.~Alboteanu, W.~Kilian and J.~Reuter,
  %``Resonances and Unitarity in Weak Boson Scattering at the LHC,''
  JHEP {\bf 0811}, 010 (2008)
  [arXiv:0806.4145 [hep-ph]].
  %%CITATION = ARXIV:0806.4145;%%

%\cite{Kunszt:1987tk}
\bibitem{Kunszt:1987tk}
  Z.~Kunszt and D.~E.~Soper,
  %``ON THE VALIDITY OF THE EFFECTIVE W APPROXIMATION,''
  Nucl.\ Phys.\  B {\bf 296} (1988) 253.
  %%CITATION = NUPHA,B296,253;%%

%\cite{Mangano:2002ea}
\bibitem{compute0l}
    F.~Caravaglios, M.~L.~Mangano, M.~Moretti and R.~Pittau,
  %``A New approach to multijet calculations in hadron collisions,''
  Nucl.\ Phys.\ B {\bf 539} (1999) 215
  [hep-ph/9807570];
  %%CITATION = HEP-PH/9807570;%%
    M.~L.~Mangano, M.~Moretti and R.~Pittau,
  %``Multijet matrix elements and shower evolution in hadronic collisions: $W b \bar{b}$ + $n$ jets as a case study,''
  Nucl.\ Phys.\ B {\bf 632} (2002) 343
  [hep-ph/0108069];
  %%CITATION = HEP-PH/0108069;%%
  M.~L.~Mangano, M.~Moretti, F.~Piccinini, R.~Pittau and A.~D.~Polosa,
  %``ALPGEN, a generator for hard multiparton processes in hadronic collisions,''
  JHEP {\bf 0307} (2003) 001
  [hep-ph/0206293];
  %%CITATION = HEP-PH/0206293;%%
 T.~Stelzer and W.~F.~Long,
  %``Automatic generation of tree level helicity amplitudes,''
  Comput.\ Phys.\ Commun.\  {\bf 81} (1994) 357
  [hep-ph/9401258];
  %%CITATION = HEP-PH/9401258;%%
    %``MadEvent: Automatic event generation with MadGraph,''
  JHEP {\bf 0302} (2003) 027
  [hep-ph/0208156].
  %%CITATION = HEP-PH/0208156;%%
  W.~Kilian, T.~Ohl and J.~Reuter,
  %``WHIZARD: Simulating Multi-Particle Processes at LHC and ILC,''
  Eur.\ Phys.\ J.\ C {\bf 71}, 1742 (2011)
  [arXiv:0708.4233 [hep-ph]].
  %%CITATION = ARXIV:0708.4233;%%
   M.~Moretti, T.~Ohl and J.~Reuter,
  %``O'Mega: An Optimizing matrix element generator,''
  In *2nd ECFA/DESY Study 1998-2001* 1981-2009
  [hep-ph/0102195].
  %%CITATION = HEP-PH/0102195;%%
  
  

%\cite{Arnold:2008rz}
\bibitem{compute1l}
  K.~Arnold, M.~Bahr, G.~Bozzi, F.~Campanario, C.~Englert, T.~Figy, N.~Greiner and C.~Hackstein {\it et al.},
  %``VBFNLO: A Parton level Monte Carlo for processes with electroweak bosons,''
  Comput.\ Phys.\ Commun.\  {\bf 180} (2009) 1661
  [arXiv:0811.4559 [hep-ph]];
  %%CITATION = ARXIV:0811.4559;%%
  K.~Arnold, J.~Bellm, G.~Bozzi, M.~Brieg, F.~Campanario, C.~Englert, B.~Feigl and J.~Frank {\it et al.},
  %``VBFNLO: A parton level Monte Carlo for processes with electroweak bosons -- Manual for Version 2.5.0,''
  arXiv:1107.4038 [hep-ph].
  %%CITATION = ARXIV:1107.4038;%%



\bibitem{FeynArts}
  T.~Hahn,
  %``Generating Feynman diagrams and amplitudes with FeynArts 3,''
  Comput.\ Phys.\ Commun.\  {\bf 140}, 418 (2001)
  [arXiv:hep-ph/0012260];
  %%CITATION = CPHCB,140,418;%%
  T.~Hahn and M.~Perez-Victoria,
  %``Automatized one-loop calculations in four and D dimensions,''
  Comput.\ Phys.\ Commun.\  {\bf 118}, 153 (1999)
  [arXiv:hep-ph/9807565];
  %%CITATION = CPHCB,118,153;%%
  T.~Hahn, M.~Rauch,
  %``News from FormCalc and LoopTools,''
  Nucl.\ Phys.\ Proc.\ Suppl.\  {\bf 157}, 236-240 (2006).
  [hep-ph/0601248].
  
\bibitem{Cuba}
  T.~Hahn,
  %``CUBA: A Library for multidimensional numerical integration,''
  Comput.\ Phys.\ Commun.\  {\bf 168}, 78-95 (2005).
  [hep-ph/0404043].



\end{thebibliography}
\end{document}